\documentclass{statsoc}

\usepackage[a4paper]{geometry}
\usepackage{graphicx,psfrag,epsf}
\usepackage[textwidth=8em,textsize=small]{todonotes}
\usepackage{enumerate}
\usepackage{natbib}
\usepackage{graphics,color}
\usepackage{verbatim}
\usepackage{setspace, subfigure, fancyhdr}
\usepackage{amsmath,amsfonts,bm,amssymb}
\usepackage{mathrsfs}
\usepackage{url}
\usepackage{arydshln}
\usepackage{tensor}
\usepackage{pdflscape}
\usepackage{ulem}
\usepackage{lineno}
\usepackage{bbm}
\usepackage{float}
\definecolor{Rcol2}{RGB}{223,83,107}
\definecolor{Rcol3}{RGB}{97,208,79}
\definecolor{Rcol4}{RGB}{34,151,230}
\definecolor{Rcol5}{RGB}{40,226,229}
\definecolor{Rcol6}{RGB}{205,11,188}
\definecolor{Rcol7}{RGB}{245,199,16}
\definecolor{Rcol8}{RGB}{158,158,158}
\definecolor{ROrange}{RGB}{255,165,0}

\title{Analyzing Cause--Specific Mortality Trends using Compositional Functional Data Analysis}
\author{Marco Stefanucci}
\address{University of Padova, Italy}
\email{marco.stefanucci@unipd.it}
\author[M. Stefanucci and S. Mazzuco]{Stefano Mazzuco}
\address{University of Padova, Italy}
\email{stefano.mazzuco@unipd.it}

\begin{document}

\newcommand{\norm}[1]{\left\lVert#1\right\rVert}

\begin{abstract}
We study the dynamics of cause--specific mortality rates among countries by considering them as compositions of functions. We develop a novel framework for such data structure, with particular attention to functional PCA. The application of this method to a subset of the WHO mortality database reveals the main modes of variation of cause--specific rates over years for men and women and enables us to perform clustering in the projected subspace. The results give many insights of the ongoing trends, only partially explained by past literature, that the considered countries are undergoing. We are also able to show the different evolution of cause of death undergone by men and women: for example, we can see that while lung cancer incidence is stabilizing for men, it is still increasing for women.
\end{abstract}


\section{Introduction}
\label{sec:intro}
Overall mortality trends may be partially explained by cause-specific data. A recent example is provided by \cite{Woolf2019} who try to shed light on the decreasing trend of US life expectancy inspecting mortality by cause, finding that midlife mortality caused by drug overdoses, alcohol abuse, suicides, and a diverse list of organ system diseases have particularly increased in the latest years. However, analyzing trends of cause specific mortality rates (CSMRs) is not straightforward, as dimensionality of data (rates by cause, year and country) might easily reach a size difficult to manage. One common solution is reducing dimensionality by collapsing one or more components: for instance, \cite{TCAL_2020} use TCAL indicator to summarize evolution over time of cause-specific mortality, but in this way it is no longer possible to explain the time trends. An additional issue when dealing with cause-specific mortality is the competing risk setting: a cause-specific mortality rate can decline because there has been a significant improvement in treatment and/or prevention of that disease or just because other causes have risen meanwhile. Therefore, if we want to analyze the time trend of CSMRs we need to take into account this feature. 
One way to do this is by means of Compositional Data Analysis (CDA) \citep{Aitchinson,pawlowsky}, which is the study of compositions, i.e. data where quantities are part of a whole, by their representation as  points on a $D$--dimensional simplex. 
Cause-specific mortality rates can be seen as compositional data, in the sense that their sum is the overall mortality rate but, if we consider their time trend, we have not only points in a simplicial space, but curves. Therefore we suggest that CDA might be combined with elements of Functional Data Analysis (FDA) \citep{FDA1}. For example, Functional PCA can be applied to reveal what are the main components driving the latest trends of CSMRs in a selected group of countries. Moreover, countries could be grouped with respect to the evolution of their CSMRs. Therefore, combining CDA and FDA we analyze trends of causes of death in 22 countries, which are eventually clustered according their CSMRs' evolution. 
The analysis proposed here is essentially descriptive, however it has the advantage of encompassing all causes of death and, at the same time, allowing to focus on specific causes.\\
The paper is organized as follows: in the next section we motivate our analysis and describe the data we used; section \ref{sec:method} describes the way in which FDA and CDA are combined together, section \ref{sec:results} shows the results and section \ref{sec:conc} illustrates the conclusions. 

\section{Motivations and Data}
\label{sec:motiv}
Cause-specific mortality rates (CSMRs) on calendar year $t$ are derived from all--causes rates $m^t_x$ and deaths for each cause $i$, age $x$ and time $t$, $\tensor*[^i]{D}{^t_x}$. Following \cite{Preston2001}, the CSMRs are defined as
\begin{equation}
  \label{eq:CSMR}
  \tensor*[^i]{m}{^t_x}= m^t_x \cdot \frac{\tensor*[^i]{D}{^t_x}}{D^t_x},
\end{equation}
where $D^t_x$ is the number of deaths for all causes occurred at time $t$ and age $x$. Therefore, considering that $\sum_i \tensor*[^i]{D}{^t_x}=D^t_x$ we have that
\begin{equation}
  \label{eq:CSMR_composition}
  \frac{\tensor*[^i]{m}{^t_x}}{m^t_x} = \frac{\tensor*[^i]{D}{^t_x}}{D^t_x} \qquad \text{and} \qquad \sum_i \frac{\tensor*[^i]{m}{^t_x}}{m^t_x} =1.
\end{equation}
 Equation \eqref{eq:CSMR_composition} shows that CSMRs are compositional data, conditional to age. Interestingly, \cite{Oeppen2} and \cite{Oeppen1} also use compositional analysis applied to cause-specific mortality, but with forecasting purposes, which lead them to consider multiple--decrement life tables deaths $\tensor*[^i]{d}{^t_x}$ rather than rates, as the sum of them over $i$ and $x$ is one (the radix of life table). So they consider $\tensor*[^i]{d}{^t_x}$ as a composition over causes $i$ and over ages $x$, and in this way they only have to make one forecasting step. However, in a descriptive perspective -- which is the one we are taking -- mixing cause pattern and age pattern, which are varying across time, countries and causes, would make interpretation more difficult, so we condition to a specific age group (40--64) that we consider more important: we excluded younger ages, as the causes involved in mortality are very limited (especially infant mortality). Ages older than 65 are also excluded, considering that the leading causes at old ages are fundamentally different from those in midlife, as also shown by \cite{Horiuchi2003}. So we focus on premature mortality cause of deaths pattern and old age can be analyzed separately.

From \eqref{eq:CSMR_composition} it follows that analyzing the ratio between cause-specific deaths ($\tensor*[^i]{D}{^t_x}$) and all-causes deaths ($D^t_x$) is equivalent to analyze the ratio between cause-specific rates ($\tensor*[^i]{m}{^t_x}$) and all-causes rate ($m^t_x$). This brings about the advantage of using a unique source of data: the \cite{WHO} mortality database, which provides data on cause and age specific deaths for all countries. Should we use also all-causes mortality rates, we would need to turn to an additional source (for instance the Human Mortality Database), but given equation \eqref{eq:CSMR_composition}, $\tensor*[^i]{D}{^t_x}$ and $D^t_x$ is all we need.

\subsection{Data}
\label{sub:data}
WHO mortality database is an archive of the causes of death information for several countries. The longest time series starts in 1950, however for many countries the information is available only since 1959. We need to consider the same time window for each country, so we focus on years 1959--2015, and we only consider countries for which data are available in this time frame. Moreover, we chose countries with relatively high longevity and quality of data, so our final analysis restricts to countries listed in table \ref{tab:countries}.
	\begin{table} 
	\caption{\label{tab:countries}Countries considered.}\\
	\centering 
	\begin{tabular}{lp{0.65\textwidth}}
	\hline
	Area					& Countries \\
	\hline
    \hline
	\textbf{North EU}		& Denmark, Finland, Norway, Sweden, Iceland \\
	\textbf{West EU}		& Austria, Belgium, Switzerland, France, Ireland, Netherlands, UK \\	
	\textbf{South EU}		& Italy, Spain, Greece \\
	\textbf{Central EU}		& Hungary, Poland \\
	\textbf{Extra-EU}		& Australia, Canada, Japan, New Zealand, USA \\
	\hline
	\end{tabular}
	\end{table}
An important issue is that classification of causes of death has greatly changed since 1959, passing from ICD $7^{th}$ to ICD $10^{th}$ revision. Following \cite{TCAL_2020}, we therefore use broad classes of causes that are little affected by changes between different revisions, see table \ref{tab:causes}. Deaths due to other causes and infant mortality, not relevant for our analysis, were removed.
\begin{table}
\caption{\label{tab:causes}Causes of death considered with related International Classification of Diseases (ICD) codes for each revision.}\\
  \centering
{\footnotesize  \begin{tabular}{|p{3.2cm}|p{3.4cm}:p{2cm}:p{2.6cm}:p{2cm}|}
\hline
Cause group                                         &ICD--7&ICD--8&ICD--9&ICD--10\\
\hline
    Certain infectious and parasitic diseases (INF) &A001--A043, A104, A132, B001--B017, B043  &A001--A044  &B01--B007 &A00-B99  \\
    \hdashline
    Endocrine, nutritional and metabolic diseases (END) &A061--A064, B020                     &A062--A066  &B18--B19  &E00-E88  \\
    \hdashline
    Circulatory system diseases (CIRC)              &A070, A079--A086, B022, B024--B029       &A080--A088  &B25--B30  &I00--I99  \\
    \hdashline
    Neoplasms (NEOP)                                &A044--A060, B018--B019                   &A045--A050, A052--A061  &B08--B09, B100, B109, B11--B17  &C00--C32, C34--D48  \\
    \hdashline
    Lung cancer (LUNG)                              &A050                                     &A051        &B101  &C33-C34  \\
    \hdashline
    Respiratory diseases (RESP)                     &A087--A097, B030--B032                   &A089--A096  &B31--B32  &J00-J98  \\
    \hdashline
    Digestive system diseases (DIG)                 &A098--A107, B033--B037                   &A097--A104  &B33--B34  &K00--K93  \\
    \hdashline
    External causes of death (EXT)                  &A138--A150, B047--B050                   &A138--A150  &B47--B56  &V00--Y89  \\
\hline
  \end{tabular}}
\end{table}
Note that some adjustments has been made: for example, HIV has been classified among Endocrine diseases in ICD--9 revision and moved to Infectious diseases class in revision 10, so we moved it in the INF class also for revision 9. In addition, we found that in Austria diabete mellitus with circulatory complications is classified among endocrine diseases, and we moved it to class CIRC (circulatory diseases) to make the classification consistent with what have been done for other countries. Then, we obtained regular curves from noisy data by spline smoothing and finally, since in some years data are missing for few countries (namely 2005 for Australia, 1997--1998 for Poland, 2000 for UK and 2015 for New Zealand), we applied the approach suggested by \cite{kraus} to impute missing parts of the curves.

\section{Methodology}
\label{sec:method}

As pointed out in the previous sections, CSMRs data can be thought as realizations of random functions taking values in the $D$--dimensional simplex and, to the best of our knowledge, there is no formal treatment of such data in the literature. \cite{muller} and \cite{menafoglio} consider the similar problem of analyzing samples of densities, i.e. functional data with the additional constraints $f(x) \geq 0$ and $\int f(x) dx=1$ that can be though as {\it continuous compositions}: the density at a specific point $x$ is an infinitesimal part of the density function and the total contribution of the parts is fixed to 1. The main difference with the present work is that in our case we consider  {\it compositional functional data}: discrete compositions where each part is a function $f_d(t)$ and the constraint $\sum_d f_d(t) =1$ holds for every $t$ on a given interval.

This section aims to furnish the main tools to analyze {\it compositional functional data} and, to do that, we extend the theory in \cite{Aitchinson} about standard compositions to deal with our complex data structure. In particular, computation of the mean function, covariance operator and principal component analysis is presented in sections \ref{simplex}-\ref{simplex2} while section \ref{clustering} discusses details about a clustering procedure we applied to data.

\subsection{The functional simplex}
\label{simplex}


A functional composition can be defined as a random function  $\bm{f}: \mathcal{I} \subset \mathbb{R} \rightarrow \mathcal{S}^D$ from a subset of the real line to the $D$--dimensional simplex. With different terms, a functional composition is a multivariate random function, $\bm{f}(t) = [f_1(t) \ldots f_D(t)]$ where each $f_d: \mathcal{I} \subset \mathbb{R} \rightarrow \mathbb{R}, d=1,\ldots, D$ is a function from a subset of the real line to $\mathbb{R}$ with the additional constraint that $f_d(t) \geq 0 \ \forall d,\forall t \in \mathcal{I} $ and $\sum_{d=1}^D f_d(t) = 1 \ \forall t \in \mathcal{I}$. We refer to realizations of functional compositions as compositional functional data. These data enjoy most of the properties of multivariate functional data but the constraint imposes some important modifications. \cite{Aitchinson} shows that compositional data lie on the simplex which has a different geometry with respect to $\mathbb{R}^D$. Similarly, compositional functional data lie in a function space that we call {\it functional simplex} for which usual operations in $\mathcal{L}_2$ cannot be applied. We define the functional simplex  $\mathcal{S}^D_f(\mathcal{I})$ as the collection of all $D$--variate functions $\bm{f}: I \subset \mathbb{R} \rightarrow \mathcal{S}^D$ that satisfy
\begin{equation}
\label{space}
 \sum_{d=1}^D \int_{\mathcal{I}} \big[\text{log}\bigg\{\frac{f_d(t)}{\tilde{f}(t)}\bigg\}\big]^2 dt < \infty  
\end{equation}
where $\tilde{f}(t) = \big[ \prod_{d=1}^D f_d(t) \big]^{1/D}$ is the geometric mean of the components of $\bm{f}$. The functional simplex is a separable Hilbert space equipped with vector operations as summation and multiplication to a scalar. Specifically, the sum of two elements in this space takes the form of a perturbation 
\begin{equation}
\label{perturbation}
\bm{f} \oplus \bm{g} = \mathscr{C}[ f_1(t) \cdot g_1(t), \ldots, f_D(t) \cdot g_D(t)],
\end{equation}
and multiplication to a scalar becomes powering
\begin{equation}
\label{powering}
\alpha \odot \bm{f} = \mathscr{C}[f_1^{\alpha}(t), \ldots, f_D^{\alpha}(t)], 
\end{equation}   
where $\mathscr{C}$ denotes the closure operator, i.e. $\mathscr{C}\bm{x} = \big[\frac{x_1}{\sum x_i}, \ldots, \frac{x_D}{\sum x_i}\big]$. These operations are the natural extensions of perturbation and powering on the simplex, and their existence is necessary for the computation of quantities relevant from a statistical point of view. In addition, the geometry of the space  can be defined by its inner product that, for any two functional compositions $\bm{f}$ and $\bm{g}$, is
\begin{equation}
\label{inner}
\langle \bm{f}, \bm{g}   \rangle_{\mathcal{S}^D_f} =  \sum_{d=1}^D \int_{\mathcal{I}} \text{log} \bigg\{\frac{f_d(t)}{\tilde{f}(t)}\bigg\} \text{log} \bigg\{ \frac{g_d(t)}{\tilde{g}(t)} \bigg\}  dt.
\end{equation} 

The centered log ratio (clr) transform \citep{Aitchinson,pawlowsky} for compositional data is a map between the $D$--dimensional simplex and $\mathbb{R}^{D-1}$. This transformation is particularly important since not only establishes an isomorphism between the two spaces, but it allows to compute quantities on the simplex through operations outside of it. Here, we define the clr transform for functional compositions. This extension can be though as the application of the standard $\text{clr}$ transform to the composition evaluated at a specific point $t$, for every $t \in \mathcal{I}$. Let $\mathcal{U}$ be the subset of the cartesian product of $D$ copies of $\mathcal{L}_2$ defined as $\mathcal{U} = \{ \bm{u}(t) \in \mathcal{L}_2^D | \sum_{d=1}^D u_d(t) = 0 \}$. The centered log ratio transform for functional compositions is a function ${\tt clr}: \mathcal{S}^D_f \rightarrow \mathcal{U} \subset \mathcal{L}_2^D$ from the functional simplex to $\mathcal{U}$ defined by 
\begin{equation}
\label{clr}
 \bm{f}^* = {\tt clr}\{\bm{f}\} = \text{log} \bigg[ \frac{f_1(t)}{\tilde{f}(t)} \ldots \frac{f_D(t)}{\tilde{f}(t)}\bigg] ,
\end{equation} 
where $\tilde{f}(t)$ is the geometric mean of the components of $\bm{f}$. This function maps elements of the functional simplex into elements of a subspace of $\mathcal{L}^D_2$, allowing us to work outside the functional simplex. To go back to the original space, we use the inverse transform ${\tt clr}^{-1}: \mathcal{U} \subset \mathcal{L}_2^D \rightarrow \mathcal{S}^D_f$ defined by
 \begin{equation}
\label{clr-1}
 \bm{f} = {\tt clr}^{-1}\{\bm{f}^*\} = \mathscr{C} \{ \text{exp} [f^*_1(t) \ldots f^*_D(t) ]\}.
\end{equation}

As an immediate consequence of these definitions we have that the functional simplex is isomorphic to $\mathcal{L}_2^{D-1}$ and the inner product defined in (\ref{inner}) is the same as the usual $\mathcal{L}_2^D$ inner product between the two clr transformed functions, i.e. $\langle \bm{f}, \bm{g}   \rangle_{\mathcal{S}^D_f} =  \langle \bm{f}^*, \bm{g}^*   \rangle_{\mathcal{L}_2^D}$. The clr transform has the important property that, for any two real constants $\alpha$ and $\beta$ and any two functional compositions $\bm{f}$ and $\bm{g}$,
\begin{equation}
\label{prop_clr}
{\tt clr} \{ (\alpha \odot \bm{f}) \oplus ( \beta \odot \bm{g})\} = \alpha \cdot {\tt clr} \{ \bm{f} \} + \beta \cdot {\tt clr} \{ \bm{g} \}.
\end{equation}

This property is central in the computation of the mean and principal components of compositional functional data, since perturbations and powerings on the functional simplex can be performed by other operations in $\mathcal{L}_2^D$.

\subsection{Mean, covariance operator and principal component analysis}  
\label{simplex2}

We define the expectation of a functional composition as $\bm{\mu}_f =\mathbb{E}\bm{f} = {\tt clr}^{-1} \{ \mathbb{E} {\tt clr} \{\bm{f}\}\} $. Suppose to have a sample of compositional functional data $\{\bm{f}_1(t), \ldots, \bm{f}_n(t) \}$ on a given time interval $\mathcal{I}$ as, for example, trends of CSMRs. The mean function $\bm{\mu}_f$ on the functional simplex representing the average trend can be estimated by
\begin{equation}
  \label{eq:mean}
   \widehat{\bm{\mu}}_f(t) = \bigoplus_{i=1}^n \bigg[ \frac{1}{n} \odot \bm{f}_i(t) \bigg].
\end{equation}

Equation \ref{eq:mean} consists in perturbations and powerings and, thanks to property  (\ref{prop_clr}) we can compute these quantities only through operations outside the functional simplex: 
\begin{equation}
  \label{eq:mean2}
   \widehat{\bm{\mu}}_f(t)  = {\tt clr}^{-1} \bigg\{ \frac{1}{n} \sum_{i=1}^n {\tt clr}\{\bm{f}_i(t)\}   \bigg\} .
\end{equation}

Note that the alternative computation of $(1/n)\sum_{i=1}^n \bm{f}_i(t)$ would lead to an improper estimator of the mean function. In the next section, we will apply estimator (\ref{eq:mean2}) to the sample of countries introduced in section \ref{sub:data} in order to depict the average trend of CSMRs.

Next we have to define a valid covariance operator for functional compositions. The definition of covariance is not straightforward for compositional data: \cite{Aitchinson} proposed three different specifications based on transformations of data. One of them relies on the clr transform and we follow this specification in our work. Basically, for a composition $\bm{x}$ each entry of the covariance matrix $\bf{R}$, $r_{jl}$, is equal to the covariance between the $j$--th and $l$--th transformed part of $\bm{x}$, i.e. $ r_{ij} = \text{cov} ({\tt clr} \{ x_j \}, {\tt clr} \{ x_l \} )$. Thus, the covariance operator of a functional composition can be defined as the integral operator $\mathscr{R} \bm{f}^*  = \int r(s,t)\bm{f}^*(t)dt$ where $r(s,t)$ is a $D \times D$ block kernel such that for block $\{ j,l\}$

\begin{equation}
\label{covariance}
 r_{jl}(s,t) = \mathbb{E} \bigg\{ {\tt clr} \{ f_j(s) \} \cdot {\tt clr} \{ f_l(t) \} \bigg\} = \mathbb{E} \bigg\{ \text{log} \bigg(\frac{f_j(s)}{\tilde{f}(s)}\bigg) \cdot \text{log} \bigg(\frac{f_l(t)}{\tilde{f}(t)}\bigg) \bigg\},
\end{equation}
where for simplicity we assume $\mathbb{E} \bm{f}^*  =0$. Similarly to the scalar case, the covariance kernel $r(s,t)$ is singular in the sense that satisfies the conditions $\sum_{j=1}^J r_{jl}(s,t) = \sum_{l=1}^L r_{jl}(s,t) =0$. However it is still positive--definite and admits the decomposition 

\begin{equation}
\label{covariance2}
 r(s,t) = \sum_{k=1}^\infty \lambda_k \bm{\phi}^*_k(s) \bm{\phi}^*_k(t), 
\end{equation}
where $\{ \lambda_1, \lambda_2, \ldots \} \in \mathbb{R}$ are the eigenvalues and $\{ \bm{\phi}^*_1, \bm{\phi}^*_2, \ldots \} \in \mathcal{U}$ are the eigenfunctions of the covariance operator $\mathscr{R}$. Since the eigenfunctions of the covariance operator $\mathscr{R}$ constitute an orthonormal basis  for $\mathcal{U}$, their compositional counterparts $\{ \bm{\phi}_1, \bm{\phi}_2, \ldots \}$ obtained through the function ${\tt clr}^{-1}$ are an orthonormal basis in the functional simplex. Finally, as a consequence of the Karhunen--Loève theorem, we can represent every functional composition as 

 \begin{equation}
  \label{eq:pca}
   \bm{f} \ominus \bm{\mu}_f = \bigoplus_{k=1}^\infty \bigg[ \xi_k \odot \bm{\phi}_k(t) \bigg],
\end{equation}
 where $  \{\bm{\phi}_1 ,  \bm{\phi}_2, \ldots \} \in \mathcal{S}^D_f$ are transformed eigenfunctions, i.e. $\bm{\phi}_k = {\tt clr}^{-1} \{ \bm{\phi}_k^*\}  \ \forall k$ and $\{ \xi_1, \xi_2, \ldots \} \in \mathbb{R}$ are computed as  $\xi_k = \langle \bm{f}, \bm{\phi}_k\rangle_{\mathcal{S}^D_f}$. Again, for (\ref{prop_clr}) we have that
 

\begin{equation}
\label{eq:pca3}
\bm{f}(t)   =  {\tt clr}^{-1} \bigg\{ {\tt clr}\{\bm{\mu}_f(t) \} + \sum_{k=1}^\infty \xi_k {\tt clr} \{\bm{\phi}_k(t) \} \bigg\}.
\end{equation}  

Given a finite sample of compositional functional data  $\{\bm{f}_1(t), \ldots, \bm{f}_n(t) \}$, the estimation of each block of the covariance kernel can be done through
\begin{equation}
\label{covariance3}
 \widehat{r}_{jl}(s,t)  = \frac{1}{n} \sum_{i=1}^n \bigg\{ \text{log} \bigg(\frac{f_{ij}(s)}{\tilde{f}_i(s)}\bigg) \cdot \text{log} \bigg(\frac{f_{il}(t)}{\tilde{f}_i(t)}\bigg) \bigg\},
\end{equation} 
where $f_{ij}(t)$ and $f_{il}(t)$ are the $j$--th and $l$--th part of the $i$--th observed function minus the sample mean. The eigenfunctions of the covariance operator $\mathscr{R}$ can be computed as the solution of the system of equations $\int r(s,t)\bm{f}^*(t) dt = \lambda \bm{f}^*(t)$, as shown in \cite{FDA1}. From a practical point of view, one has to compute empirical eigenfunctions $\{ \widehat{\bm{\phi}}_1^*,\widehat{\bm{\phi}}_2^*, \ldots \}$ using the covariance kernel $\widehat{r}(s,t)$ estimated from data, solving the problem  

\begin{equation}
\label{eigen}
 \int \widehat{r}(s,t)\bm{f}^*(t) dt = \lambda \bm{f}^*(t),
\end{equation}
then, transformed eigenfunctions can be obtained as $\widehat{\bm{\phi}}_k = {\tt clr}^{-1} \{ \widehat{\bm{\phi}}_k^* \}$. 

Computation of the empirical eigenfunctions is important because the result provides information about the main modes of variation in the considered sample. For complex data as CSMRs over time, these functions represents a straightworward way to inspect the simultaneous directions of variability of all causes of death. The empirical scores $\{ \widehat{\xi}_{i1}, \widehat{\xi}_{i2},\ldots\}$ for the $i$--th observation $\bm{f}_i$ can be computed as 

\begin{equation}
\label{eq:scores}
  \widehat{\xi}_{ik} = \langle \bm{f}^*_i, \bm{\phi}_k^* \rangle = \int {\tt clr}\{ \bm{f}_i\}(t) \bm{\phi}_k^*(t) dt .
\end{equation}

Finally, by truncating the infinite series in (\ref{eq:pca3}) and replacing population quantities with estimates, one obtains the approximation

\begin{equation}
\label{eq:recon}
    \bm{f}(t) \simeq  {\tt clr}^{-1} \bigg\{ {\tt clr}\{\widehat{\bm{\mu}}_f(t)  \} + \sum_{k=1}^K \widehat{\xi}_k {\tt clr} \{\widehat{\bm{\phi}}_k(t) \} \bigg\}.
\end{equation}  

This representation is particularly useful because it gives an approximation of a functional composition using few nonrandom functions shared among observations and few scalar random variables, thus reducing the problem to a finite--dimensional one. The relevance in our application is that CSMRs observations can be summarized by their behaviour with respect to few main modes of variation. The projections on the finite--dimensional space can also serve as a basis for a clustering procedure, as explained in the next subsection.

\subsection{The Clustering Step}
\label{clustering}

A common practice in FDA is to reduce the dimensionality of data and use projections on a finite dimensional space as a starting point for other procedures, such as clustering or classification (see for instance \cite{muller2,sangalli}). In our context, the compositional functional PCA allows us to drastically reduce the complexity of CSMRs over time by considering only few components, accounting for most of the variability in the sample. Then, in order to detect a clustering structure, one possibility is to apply a procedure for euclidean data to the scores of our original compositional functional data. 

Spectral clustering (see \cite{spectral} for an overview) is a relatively recent clustering method. It is based on the eigenanalysis of the Laplacian matrix constructed on the similarity graph of the data. Its advantage over alternative procedures is the ability to detect complex and nonlinear structures. The application of this method to our context needs no further adjustments with respect to the original formulation, but we have to choose some parameters. In particular, we set the gaussian similarity kernel defined as $s(\bm{\xi}_i, \bm{\xi}_j) =  \text{exp}(-\norm{(\bm{\xi}_i-\bm{\xi}_j}^2/(2\sigma^2)) $
with $\sigma =1$ as kernel to compute the similarity graph among scores, and k--means \citep{multivariate} as base method. The output of the clustering procedure is given by a set of $G$ centroids $\bm{\gamma}_1, \ldots, \bm{\gamma}_G$ and a vector of labels of cluster membership. Since the method is sensible to initialization, we run spectral clustering algorithm for $B$ times and compute the {\it majority vote} as the most frequent partition of data over $B$ repetitions. For this method, the decay of the spectrum of the Laplacian matrix can be informative about the number of clusters. However, we did not observe a clear signal when we applied the method to CSMRs data. Thus we decided to use the silhouette index for determining the number of clusters, defined as the average of all silhouette values $s(i)$, 

\begin{equation}
\label{silhouette}
s(i) = \frac{b(i)-a(i)}{\text{max} \{ a(i),b(i) \}} ,
\end{equation}
\begin{equation}
\label{silhouette2}
 b(i) = \underset{k \neq i}{\text{min}}\frac{1}{|C_k|} \sum_{j \in C_k} (\bm{\xi}_i - \bm{\xi}_k )^2, \quad \quad \quad a(i) = \frac{1}{|C_i|-1} \sum_{j \in C_i, i \neq j}(\bm{\xi}_i - \bm{\xi}_j )^2, 
\end{equation}

where $\bm{\xi}_i$ is the vector of scores for the $i$--th observation and $|C_k|$ is the cardinality of the $k$--th cluster \citep{multivariate}. The silhouette index ranges from -1 to 1 and higher values correspond to better cluster configurations.

\section{Results}
\label{sec:results}

In this section we present the results of the analysis on the WHO mortality database separately for men and women. We first compute the compositional functional mean as in (\ref{eq:mean2}), then we estimate the covariance kernel using (\ref{covariance3}) and compute principal components and scores with (\ref{eigen}) and (\ref{eq:scores}). Finally, we apply spectral clustering to the scores in order to obtain a grouping structure of the countries in the study. The silhouette values for each country are reported in table 1 of the supplementary material.

\subsection{Men}

\begin{figure}[H]
\includegraphics[width=\linewidth]{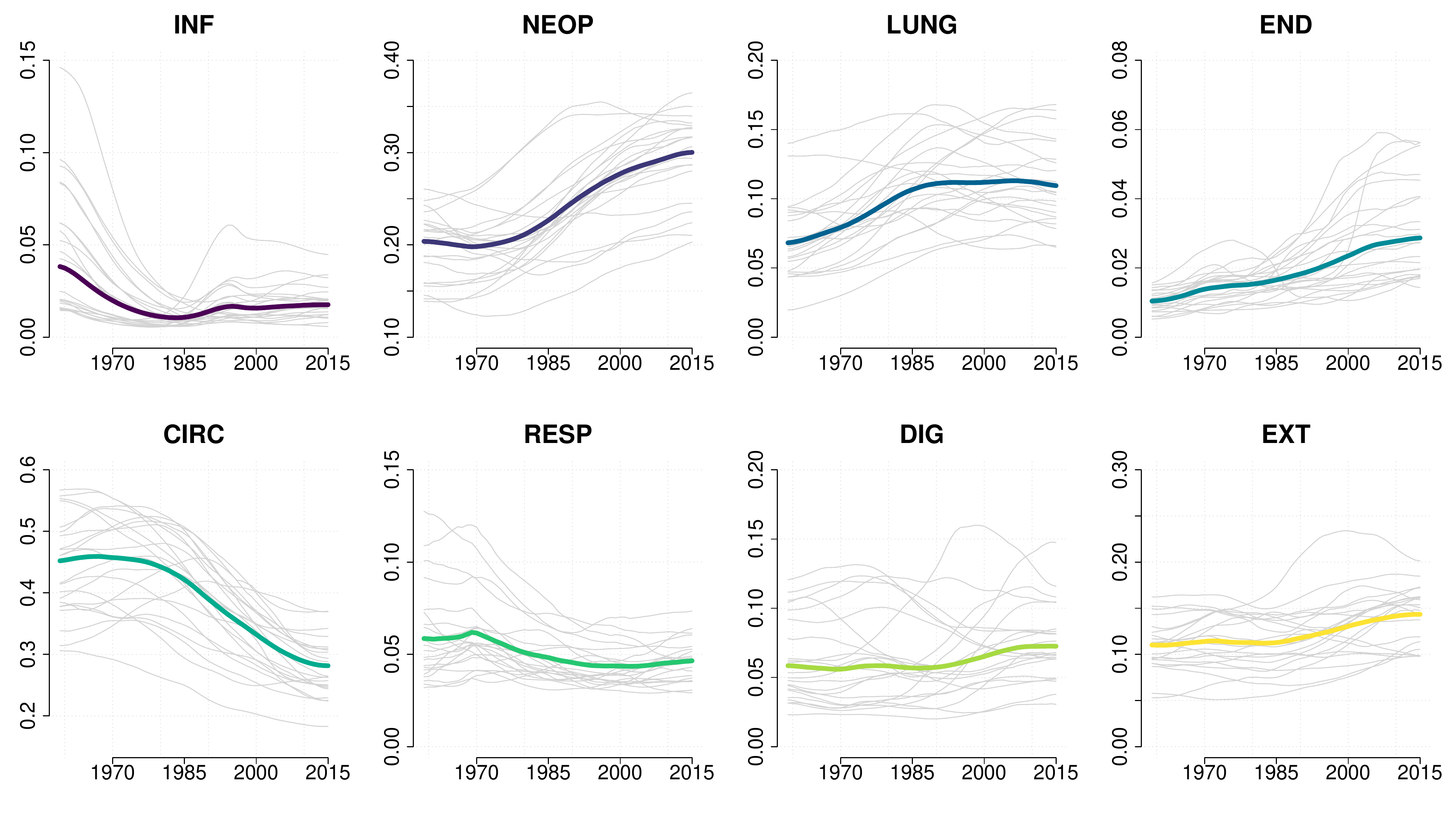}
\caption{Composition of mortality rates over years for men population. Curves for 22 countries in each panel are colored in grey. Colored curves represent the compositional functional mean.}
\label{causes-men}
\end{figure}

The 22 curves of the men sample for 8 causes of death we consider are given in figure \ref{causes-men}. We depict with a larger colored line the compositional functional mean. The ranges among causes are different, denoting a different contribution on the total mortality rate. Highest ranges are  
\noindent
for circulatory diseases (0.20--0.50) and neoplasms (0.10--0.40), representing the two most important causes of death from 1959 to 2015. However, dynamics are quite different among all causes: neoplasms and lung cancer in particular exhibit an increasing trend over years, while circulatory diseases constantly decrease their contribution. 
A slightly increasing trend can be noticed also for endocrine, metabolic and nutritional diseases and external causes while respiratory and digestive diseases show heterogeneous patterns: some countries have a declining trend, others an increasing one. Infectious diseases experienced a particular behavior over years: they diminish their contribution at the beginning of the period, then started to increase and finally stabilized around a value of 0.02. 
These results are in line with the epidemiological transition theory with increasing prevalence of diseases associated with aging (e.g., neoplasms) and of man-made diseases (lung cancer caused by smoking, metabolic and nutritional disorders, caused by obesity and external causes of deaths) and a decline of infectious diseases.

\begin{figure}[t]
\includegraphics[width=\linewidth]{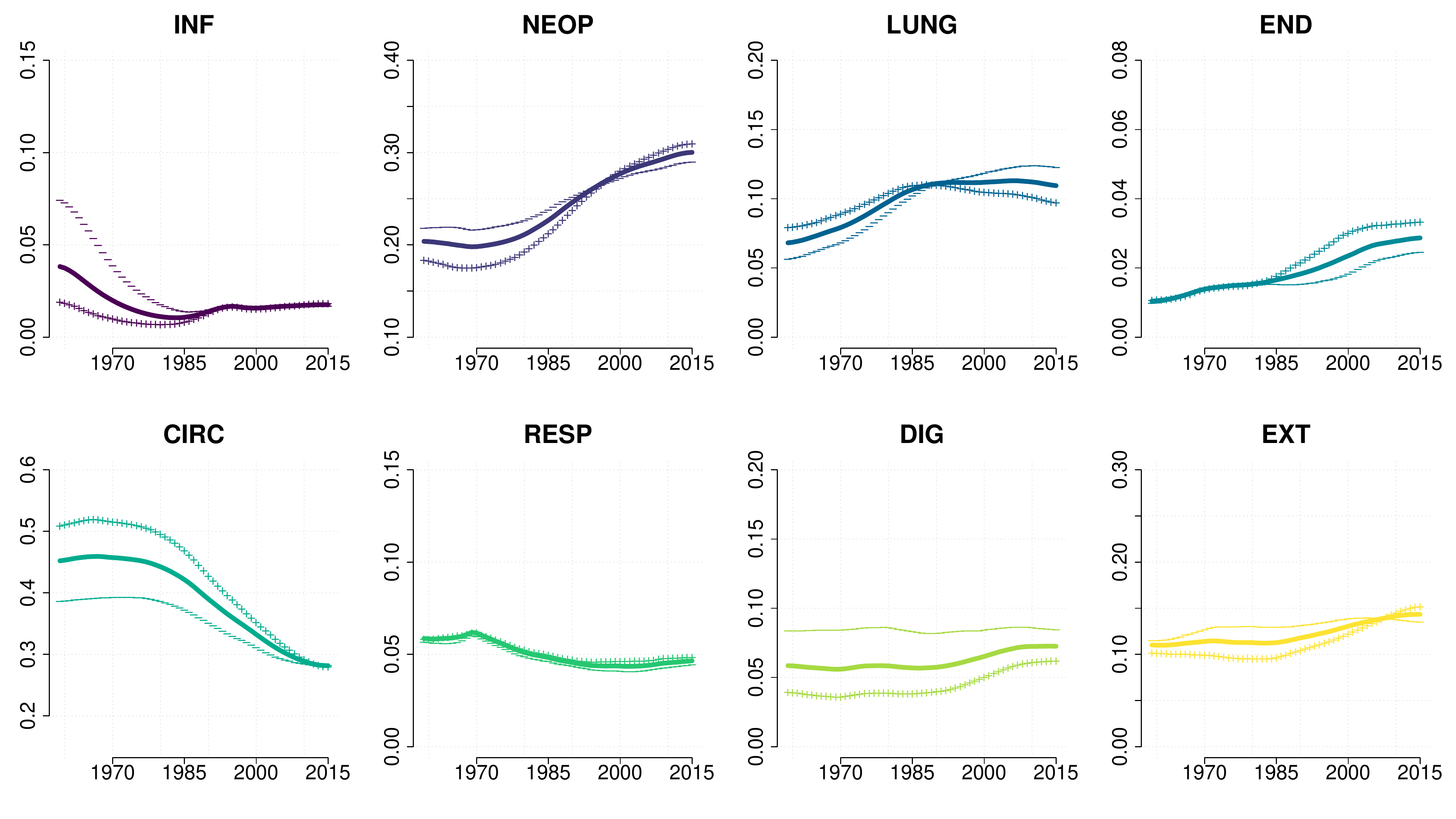}
\caption{First compositional functional principal component for men population. The continuous line represents the mean function while outer lines represent the mean function +/- the component. }
\label{PC1-men}
\end{figure}

The first compositional functional principal component for men population is depicted in figure \ref{PC1-men}. This function is computed as described in section \ref{simplex2} and represents the main mode  of variation of the men sample. The eigenvalue associated with this eigenfunction is $\lambda_1 = 17.06$ and the Fraction of Explained Variance (FEV) is $\lambda_1/\sum \lambda_k = 0.337$, thus this component account for about one third of the total variability. This component is related to all causes but respiratory diseases. A variability connected with the {\it level} of the cause -- values always above or below the mean -- is present in digestive diseases. For infectious and endocrine diseases, as we already noticed, the main variability is related only to specific years -- 1959-1985 for the former and 1985-2015 for the latter -- and we call this aspect {\it local} level variability. This behaviour can be partially observed for circulatory diseases and external causes too. Lastly, a more structured 

\begin{figure}[H]
\includegraphics[width=\linewidth]{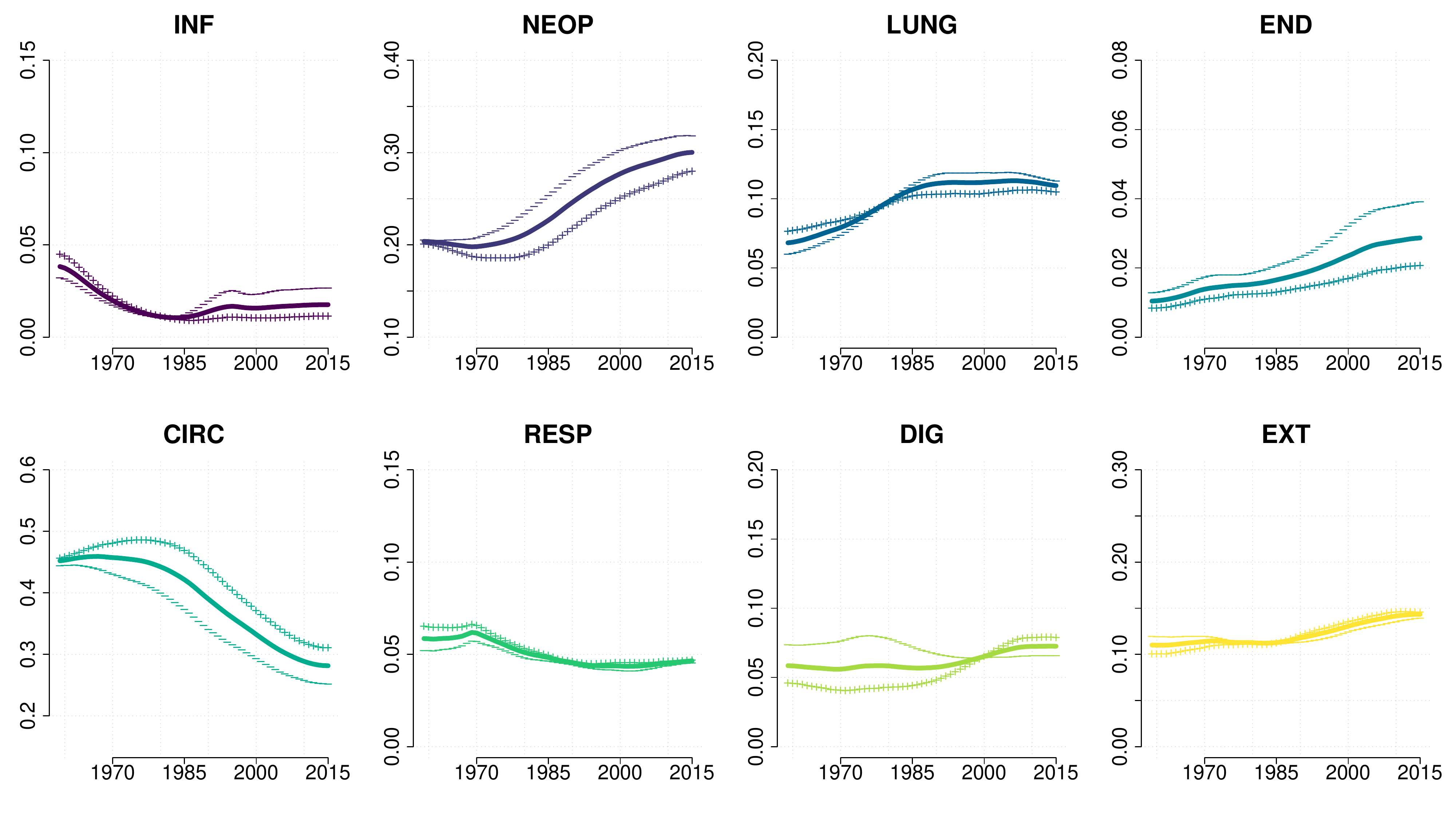}
\caption{Second compositional functional principal component for men population. The continuous line represents the mean function while outer lines represent the mean function +/- the component. }
\label{PC2-men}
\end{figure}
\noindent
pattern can be seen for neoplasms and lung cancer: for these causes the first compositional functional principal component crosses the mean at some point. This reflects the fact that in 
 \begin{figure}[ht]
\includegraphics[width=\linewidth]{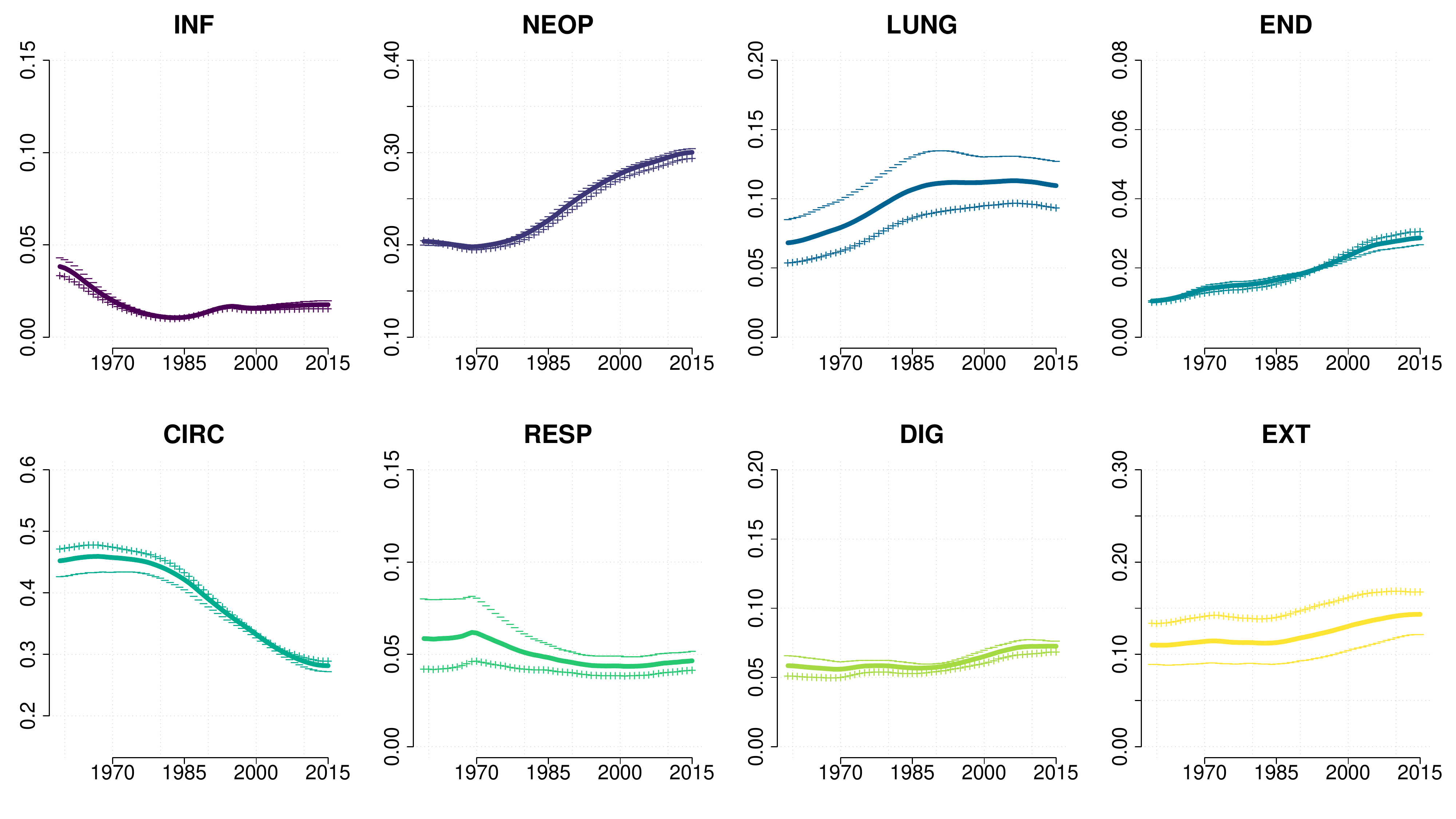}
\caption{Third compositional functional principal component for men population. The continuous line represents the mean function while outer lines represent the mean function +/- the component. }
\label{PC3-men}
\end{figure}

\begin{figure}[H]
\includegraphics[width=\linewidth]{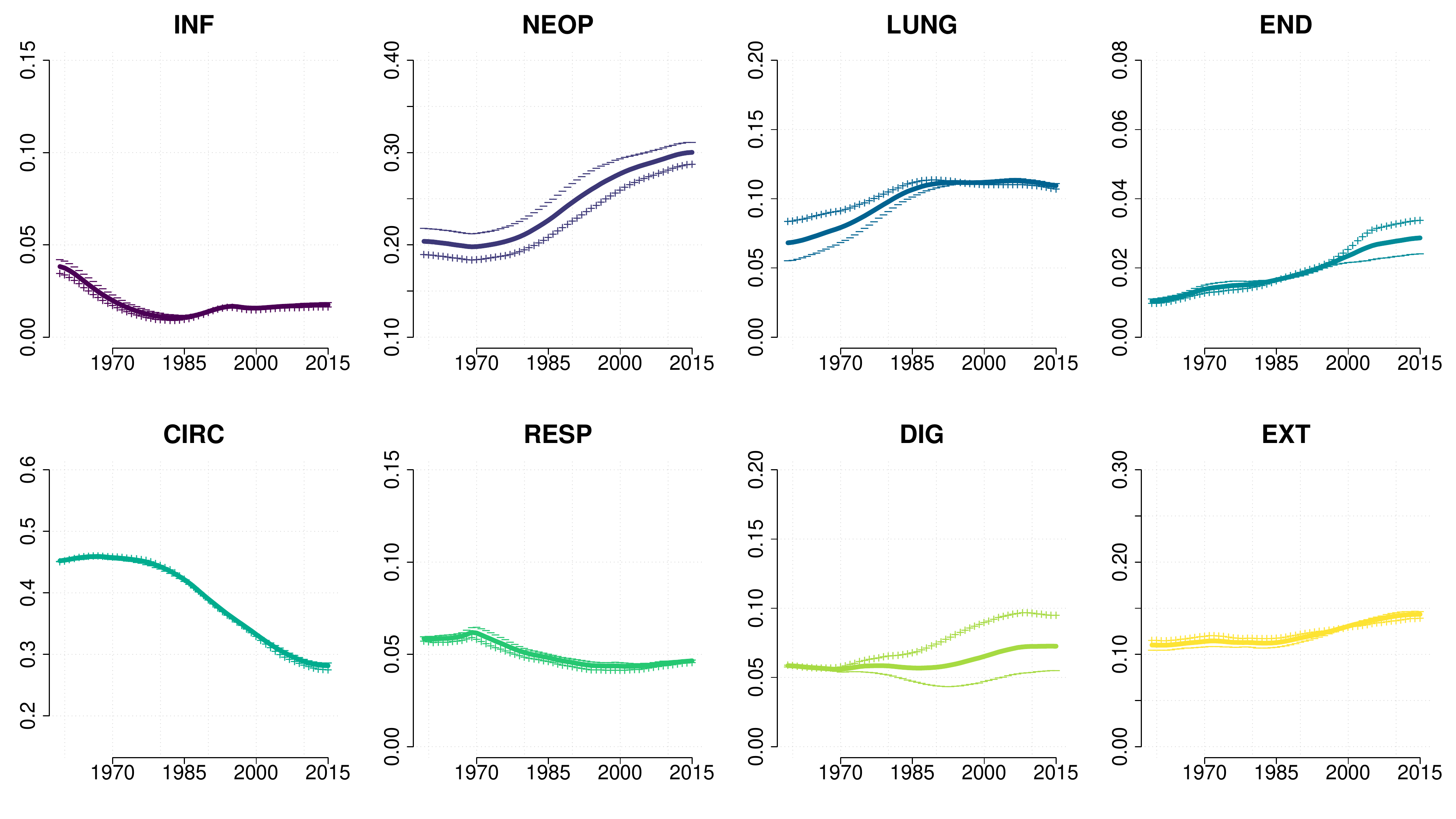}
\caption{Fourth compositional functional principal component for men population. The continuous line represents the mean function while outer lines represent the mean function +/- the component. }
\label{PC4-men}
\end{figure}
\noindent
our sample some countries had a faster increase of this causes with respect to the mean while others had a slower increase than the mean.  
Overall, a positive score on this component would lead to a high level of circulatory diseases and endocrine diseases (only after 1985), a low level of digestive diseases and infectious diseases (only before 1985), a fast increase of neoplasms and a slow increase of lung cancer. Negative scores characterize an opposite behaviour. The scores obtained by each country on the first 4 PCs for men sample are reported in table \ref{tab:scores}.

Figure \ref{PC2-men} shows the second compositional functional principal component for men population. The eigenvalue associated with this eigenfunction is $\lambda_2 = 10.92$ and the FEV is $0.216$. This component represents a {\it level} variability for neoplasms, circulatory and endocrine diseases, especially in the last 30 years. {\it Local} variability is observed for infectious diseases -- note that the time interval connected with this component is 1985-2015 -- and for respiratory diseases, but to a lesser extent. For lung cancer, digestive diseases and external causes this component shows a cross--mean behaviour, reflecting different velocities among countries in the evolution of these contributions. Overall, a positive score on this component would lead to a high level of circulatory diseases, a low level of neoplasms and endocrine diseases, a fast decay of infectious diseases, a slow increase of lung cancer and an increase of digestive diseases after 2000.

The third compositional functional principal component for men population is illustrated in figure \ref{PC3-men}. The eigenvalue associated with this eigenfunction is $\lambda_3 = 7.27$ and the FEV is $0.144$. This component is mostly connected with four causes of death. It represents a {\it level} variability for lung cancer, respiratory diseases and external causes while {\it local} variability is observed for circulatory diseases, in relation to the period 1959-1990. 
Overall, a positive score on this component would lead to a low level of lung cancer and respiratory diseases and to a high level of external causes and circuatory diseases -- especially at the beginning of the considered period. 

 \begin{table} 
  \caption{\label{tab:scores} Scores of the 22 countries in the study. Men sample is reported on the left and women sample on the right.}\\
  \centering
  \begin{minipage}{.5\textwidth}
  \centering 
  \begin{tabular}{lrrrr}
  \hline
	Countries					& PC1 & PC2 & PC3 & PC4\\
	\hline
 \hline
AUS		& 4.38 & -2.08 & 0.32 & 0.28 \\
AUT		& -3.83 & -1.03 & 0.95 & 3.90 \\
BEL		& 0.44 & 1.43 & -3.02 & -0.21 \\
CAN		& 3.26 & -3.80 & 0.80 & 0.75 \\
DNK		& 3.41 & -1.51 & 0.23 & 3.42 \\
FIN		& -0.63 & 6.34 & 3.59 & 3.36 \\
FRA		& -5.20 & -3.66 & -0.10 & 0.18 \\
GRE		& -2.90 & 1.77 & -2.14 & -2.28 \\
HUN		& -5.80 & 4.11 & 0.09 & 3.32 \\
ICE		& 3.01 & 3.13 & 4.77 & -3.32 \\
IRL		& 2.15 & 4.72 & -2.67 & -2.63 \\
ITA		& -2.49 & -3.81 & -2.90 & 0.64 \\
JPN		& -6.76 & -2.10 & 2.89 & -2.98 \\
NL		& 5.55 & -1.00 & -2.77 & 0.72 \\
NZL		& 6.49 & -0.69 & 0.82 & -2.21 \\
NOR		& 2.47 & 0.72 & 2.93 & -1.17 \\
POL		& -5.27 & 3.93 & 0.19 & -1.58 \\
SPA		& -5.76 & -2.51 & -3.61 & -2.02 \\
SWE		& 1.37 & -1.45 & 4.52 & 0.43 \\
SWI		& -0.65 & -2.19 & 0.55 & 0.06 \\
UK		& 4.28 & 4.84 & -5.20 & 1.05 \\
USA		& 2.50 & -5.16 & -0.24 & 0.27 \\

	\hline
	\end{tabular}
	\end{minipage}
	\begin{minipage}{.5\textwidth}
	\centering 
	\begin{tabular}{lrrrr}
	\hline
	Countries					& PC1 & PC2 & PC3 & PC4\\
	\hline
\hline
AUS		& 3.08 & 1.87 & 0.86 & -0.42 \\
AUT		& -2.76 & 1.02 & -1.31 & -3.53 \\
BEL		& -1.67 & 2.06 & -0.51 & -0.01 \\
CAN		& 2.74 & 3.76 & 0.17 & -0.37 \\
DNK		& 4.12 & 1.92 & -2.26 & -3.47 \\
FIN		& -3.01 & -2.79 & -3.92 & 0.06 \\
FRA		& -6.77 & 2.70 & -2.35 & 0.90 \\
GRE		& -0.52 & -2.37 & 2.50 & 0.72 \\
HUN		& -2.96 & -3.25 & -1.79 & -5.40 \\
ICE		& 5.40 & 0.21 & -4.05 & 4.43 \\
IRL		& 4.00 & -5.82 & 1.25 & 1.01 \\
ITA		& -3.72 & 1.75 & 4.19 & -1.47 \\
JPN		& -4.21 & -1.69 & -0.99 & 4.99 \\
NL		& 1.98 & 2.59 & 0.75 & -1.22 \\
NZL		& 5.10 & -0.62 & 3.00 & 0.59 \\
NOR		& 2.37 & -0.71 & -1.49 & 1.85 \\
POL		& -3.63 & -4.13 & 1.14 & -0.74 \\
SPA		& -5.71 & -0.81 & 4.00 & 1.87 \\
SWE		& 1.02 & 1.40 & -2.02 & 0.17 \\
SWI		& -2.54 & 2.31 & -1.56 & 0.81 \\
UK		& 5.97 & -3.21 & 0.61 & -2.05 \\
USA		& 1.73 & 3.82 & 3.76 & 1.29 \\

	\hline
	\end{tabular}
	\end{minipage}
	\end{table}

Figure \ref{PC4-men} depicts the fourth compositional functional principal component for men population. The eigenvalue associated with this eigenfunction is $\lambda_4 = 4.62$ and the FEV is $0.091$. Also this component refers to only four causes of death. It represents a {\it level} variability for neoplasms and a {\it local} variaility for lung cancer (1959-1995), endocrine diseases (after 2000) and digestive diseases (from 1970). Overall, a positive score on this component would lead to a low level of neoplasms and a high level of lung cancer, endocrine diseases and digestive diseases in the aforementioned periods. 
 
\begin{figure}[t]
\includegraphics[width=.45\linewidth]{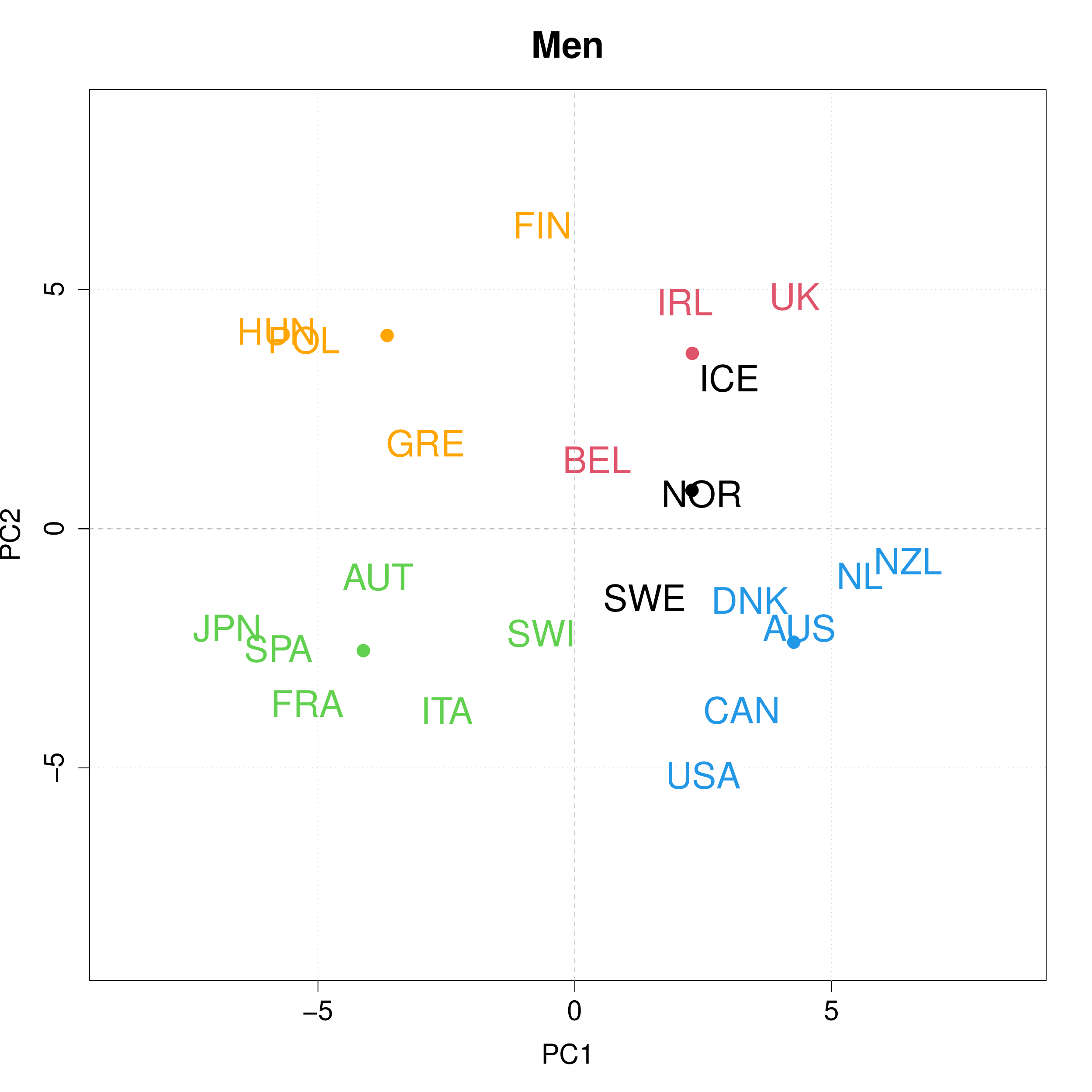}\hfill
\includegraphics[width=.45\linewidth]{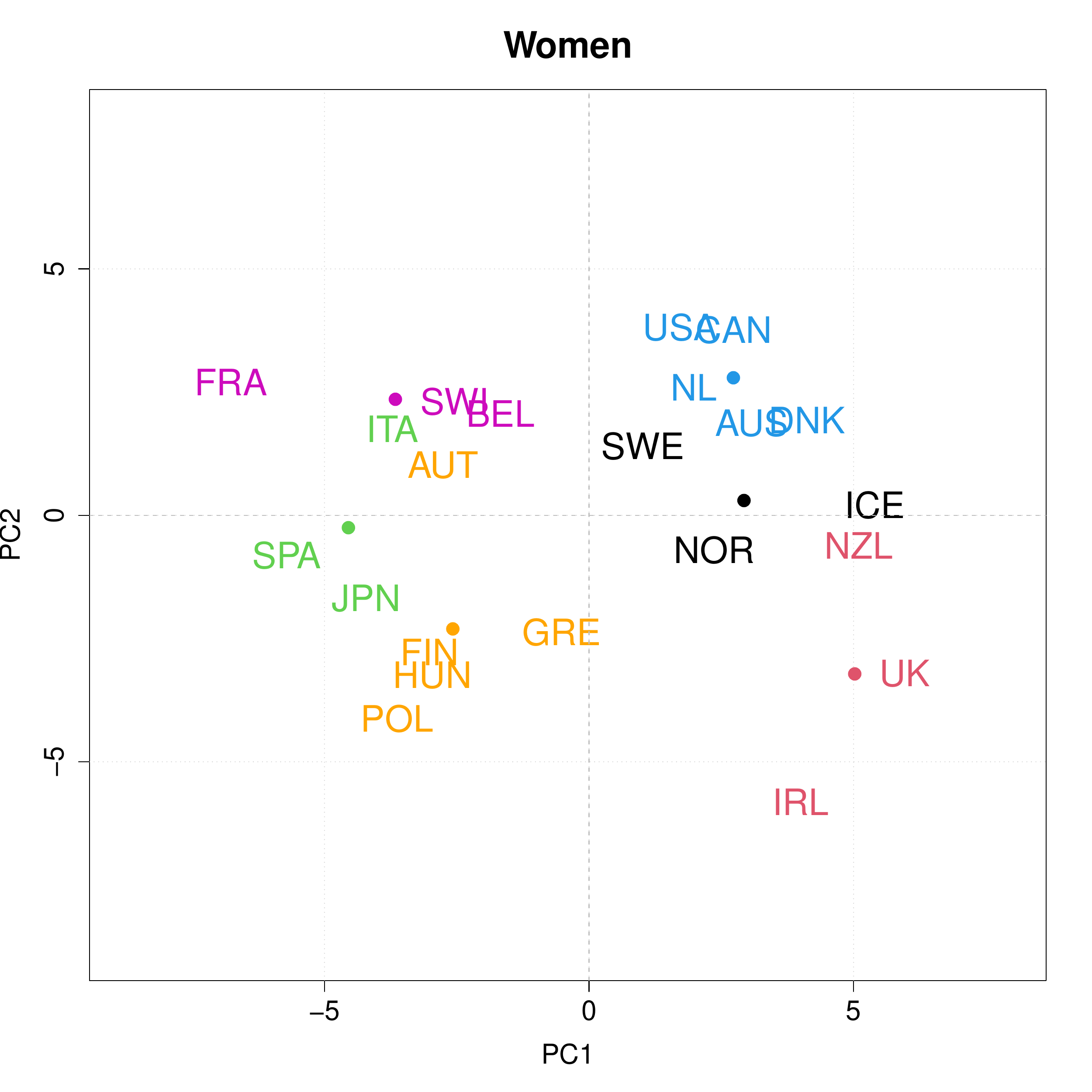}
\caption{Scores $\xi_1$ and $\xi_2$ for countries in the sample. Different colors represent different clusters.}
\label{scores}
\end{figure}

\begin{figure}[t]
\includegraphics[width=\linewidth]{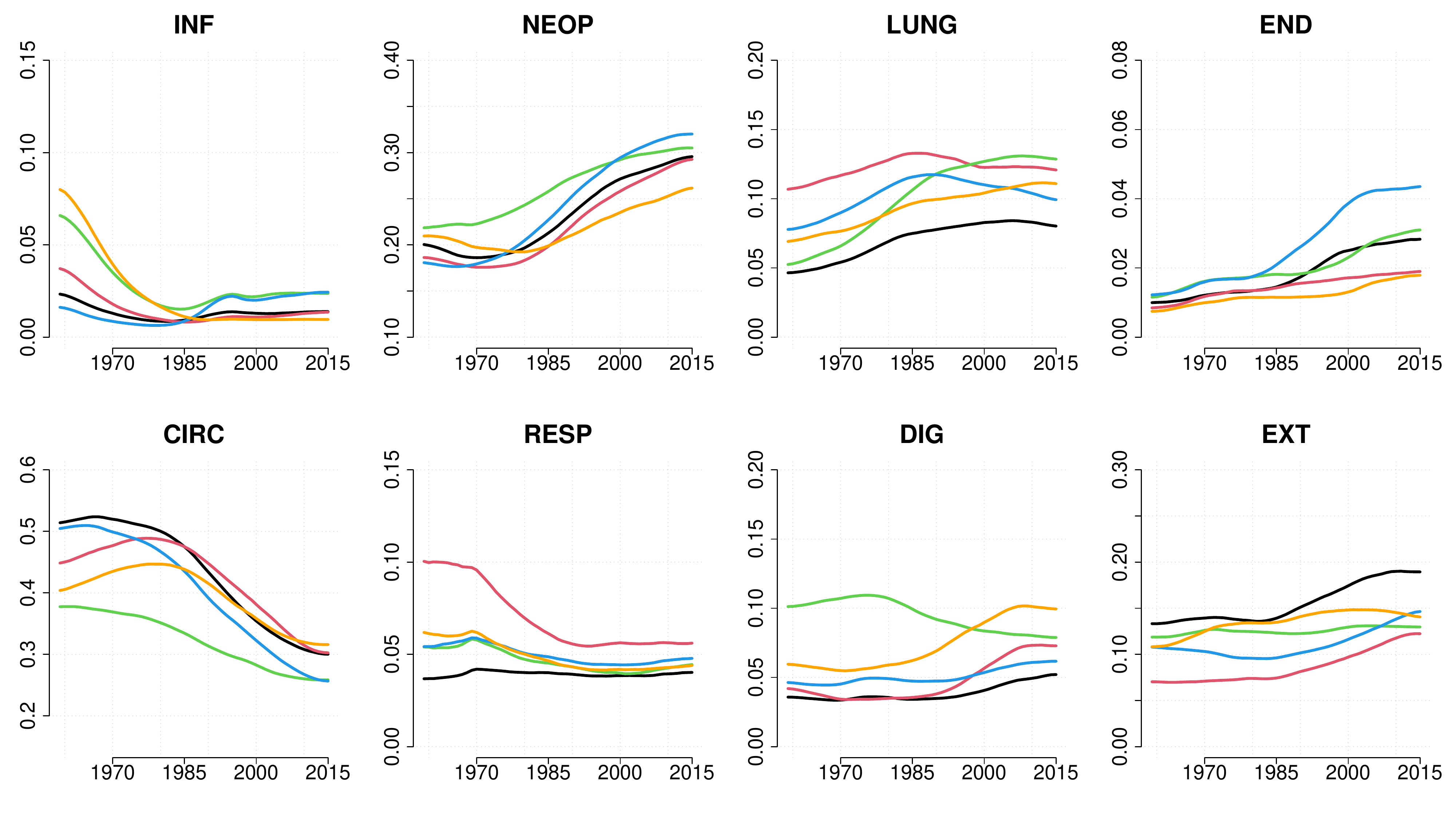}
\caption{Functional centroids of the spectral clustering for men sample. Legend: {\color{Rcol4} \textbf{----}} Cluster 1, {\color{Rcol3} \textbf{----}} Cluster 2, {\color{ROrange} \textbf{----}} Cluster 3, {\color{Rcol2} \textbf{----}} Cluster 4,  \textbf{----} Cluster 5.}
\label{centroids-men}
\end{figure}

A clustering procedure has been applied to the men sample as described in section \ref{clustering}. We considered the first $4$ PCs, accounting for 78.7\% of the total variability. For different values of $G$, we ran the spectral clustering algorithm for $B=1000$ times and save the result of the {\it majority vote} . The spectrum of the Laplacian matrix shows no evidence for a particular number of clusters, while the silhouette index is maximized for $G=5$ clusters, which are reported in table \ref{tab:clus-men}. A visual representation of the scores of the first two PCs along with the clustering structure can be found in figure \ref{scores}. Using (\ref{eq:recon}) with $K=4$ and $\gamma_{1g},\gamma_{2g},\gamma_{3g},\gamma_{4g}$ as the centroids of the spectral clustering output, we reconstructed the compositional functional centroids $\bm{f}_g, g=1,\ldots,G$, depicted in figure \ref{centroids-men}. These trajectories summarize the behavior of each cluster with respect to the eight causes of death.

We can draw some comments. The first group (including USA, Canada, Australia, New Zealand, Netherlands and Denmark) is characterized by relatively low circulatory related deaths (but at the beginning of the observation period it was high) a decreasing trend of lung cancer and an almost plateauing trend of neoplasms. Notably the level of endocrine and metabolic related deaths is much higher than other clusters, with a particularly high slope between 1980 and 2000.
This trend is close to that of obesity prevalence in USA, as shown, for example by \cite{USA_obesity}, and New Zealand and Australia show similar figures, so it is likely that this is the leading cause. It should also be noted that external causes of death are rapidly increasing in the last years, in line with what is shown by \cite{Woolf2019}. The second group consists of Italy, Spain, France, Japan, Austria and Switzerland, and is characterized by high lung cancer mortality -- in all countries smoking prevalence is high among men, see for instance,
 	\begin{table}
	\caption{\label{tab:clus-men}Clustering output for men sample.}\\
	\centering 
	\begin{tabular}{lp{0.65\textwidth}}
	\hline
	Clusters					& Countries \\
	\hline
\hline
	\textbf{Cluster 1}		& USA, Canada, Australia, New Zealand, Denmark, Netherlands  \\
	\textbf{Cluster 2}		& Italy, France, Spain, Austria, Switzerland, Japan  \\	
	\textbf{Cluster 3}		& Hungary, Poland, Finland, Greece\\
	\textbf{Cluster 4}		&  UK, Ireland, Belgium\\
	\textbf{Cluster 5}		&  Norway, Sweden, Iceland\\
	\hline
	\end{tabular}
	\end{table}
	\begin{table} 
	\caption{	\label{tab:clus-women}Clustering output for women sample.}\\
	\centering 
	\begin{tabular}{lp{0.65\textwidth}}
	\hline
	Clusters					& Countries \\
	\hline
\hline
	\textbf{Cluster 1}		& USA, Canada, Australia, Denmark, Netherlands  \\
	\textbf{Cluster 2}		& Italy, Spain,  Japan\\
	\textbf{Cluster 3}		& France, Switzerland, Belgium\\
	\textbf{Cluster 4}		& Hungary, Poland, Austria, Finland, Greece\\
	\textbf{Cluster 5}		&  UK, Ireland, New Zealand\\
	\textbf{Cluster 6}		&  Norway, Sweden, Iceland\\
	\hline
	\end{tabular}
	\end{table}
\cite{smoke1} -- and low circulatory-related deaths. At the beginning of the observation period up to 1980, this group had particularly high level of digestive system diseases and infectious related deaths, as for most of them the epidemiological transition started later than in other countries. Third group is made up of Hungary, Poland, Finland and Greece. The main feature of this group is the increasing relevance of digestive system diseases related deaths. This might be driven by rising alcohol--attributable diseases, considering that such health problems are a growing concern in these countries. It must be noted that this is the least homogeneous group, as can be seen from table 1 of the supplementary material. Finland, for example, has a peculiar pattern with an extremely high incidence of external causes of deaths that is more consistent with the group including Sweden, Norway and Iceland  but, at the same time, a rapidly increasing prevalence of digestive diseases that matches more with Eastern Europe group. Greece, instead is in between cluster 3 and 2 having a particularly high incidence of lung cancer (as in cluster 2) but also a high level -- and very slowly declining -- of circulatory diseases related deaths, as in cluster 3. The fourth group is made up of UK, Ireland and Belgium, characterized by high -- albeit descending -- levels of lung cancer, respiratory diseases and circulatory diseases and increasing relevance of digestive system diseases. This might be brought about by risk factors like smoking, alcohol consumption and poor diet. The last cluster includes Sweden, Norway and Iceland, characterized by the lowest level of lung cancer and respiratory diseases and the highest level of external causes.

\subsection{Women}

\begin{figure}[H]
\includegraphics[width=\linewidth]{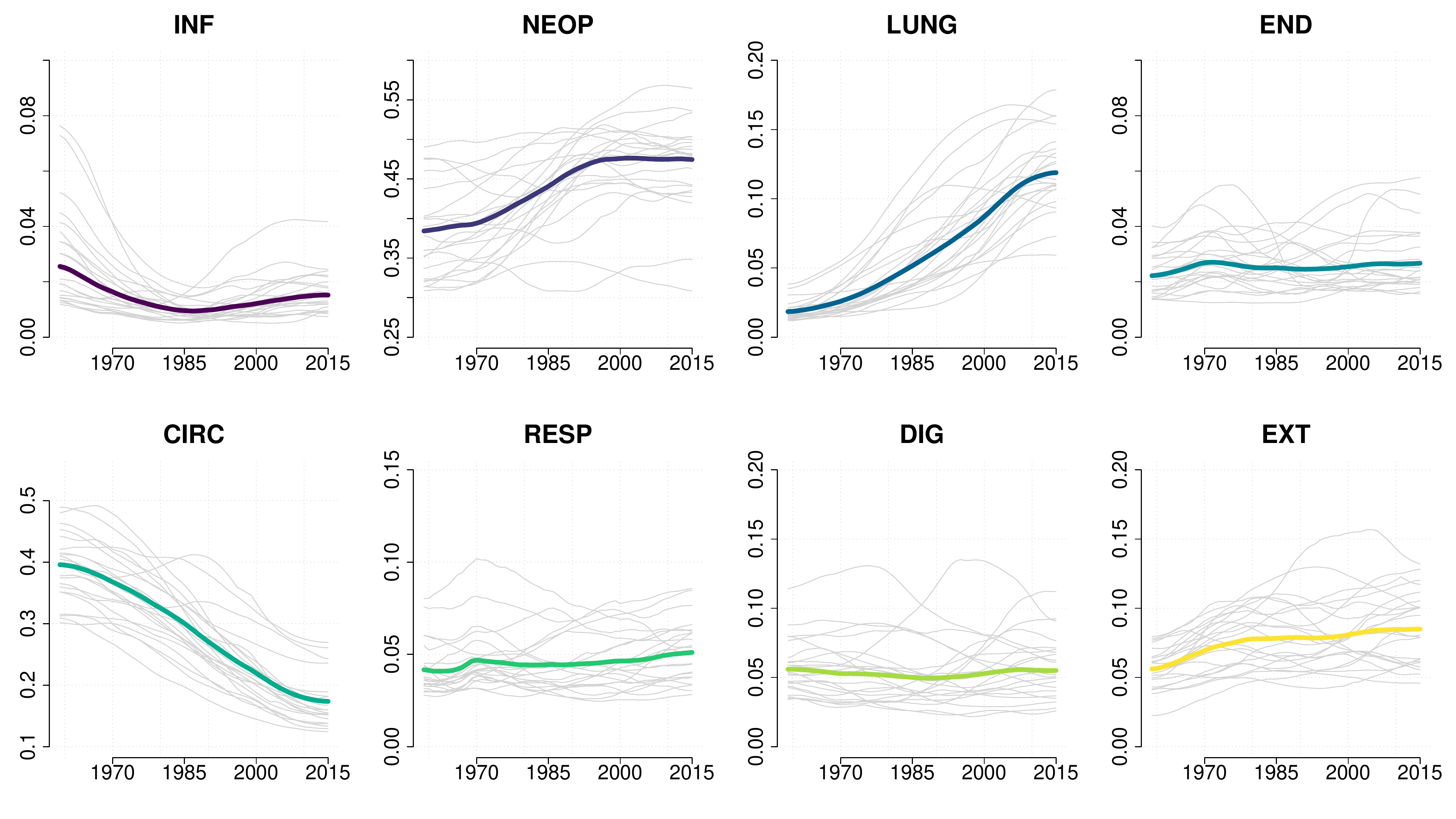}
\caption{Composition of mortality rates over years for women population. Curves for 22 countries in each panel are colored in grey. Colored curves represent the compositional functional mean.}
\label{causes-women}
\end{figure}

The 22 curves of the women sample for 8 causes of death we consider are represented in figure \ref{causes-women}.
Similarly to men population, the ranges among causes are different. Highest ranges are observed for circulatory diseases (0.1--0.4), neoplasms (0.3--0.6) and lung cancer (0.02--0.2). With respect to men, women have a higher contribution of neoplasms and a lower contribution of circulatory diseases, though trends over years are similar: the former moved from 0.38 to 0.48 and the latter reduced from 0.4 to below 0.2. Lung cancer had almost negligible contribution at the beginning of the considered period but, thanks to a strong and constant increase, it is nowadays as important as in men (slight above 0.1). On the other side, the positive trend for other neoplasms seems to have reached a plateau since the last decade of the twentieth century. An opposite behaviour can be found in men, where lung cancer had a stable contribution since 1985 while other neoplasms are still increasing. Another difference can be found for endocrine diseases, where women do not show an increasing trend. Lastly, infectious diseases experienced a similar evolution for both men and women. 
%
%

\begin{figure}[t]
\includegraphics[width=\linewidth]{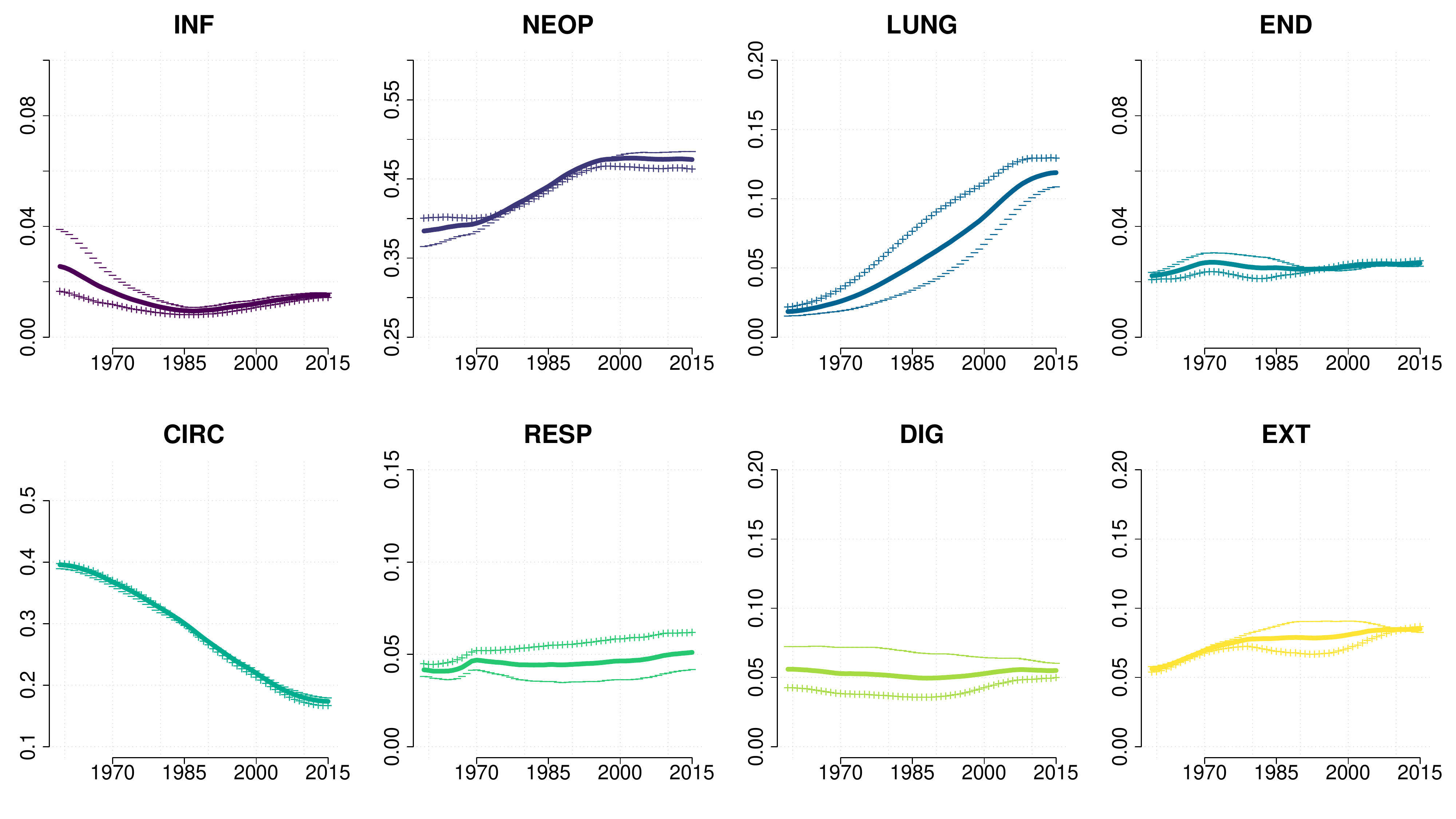}
\caption{First compositional functional principal component for women population. The continuous line represents the mean function while outer lines represent the mean function +/- the component. }
\label{PC1-women}
\end{figure}

\begin{figure}[H]
\includegraphics[width=\linewidth]{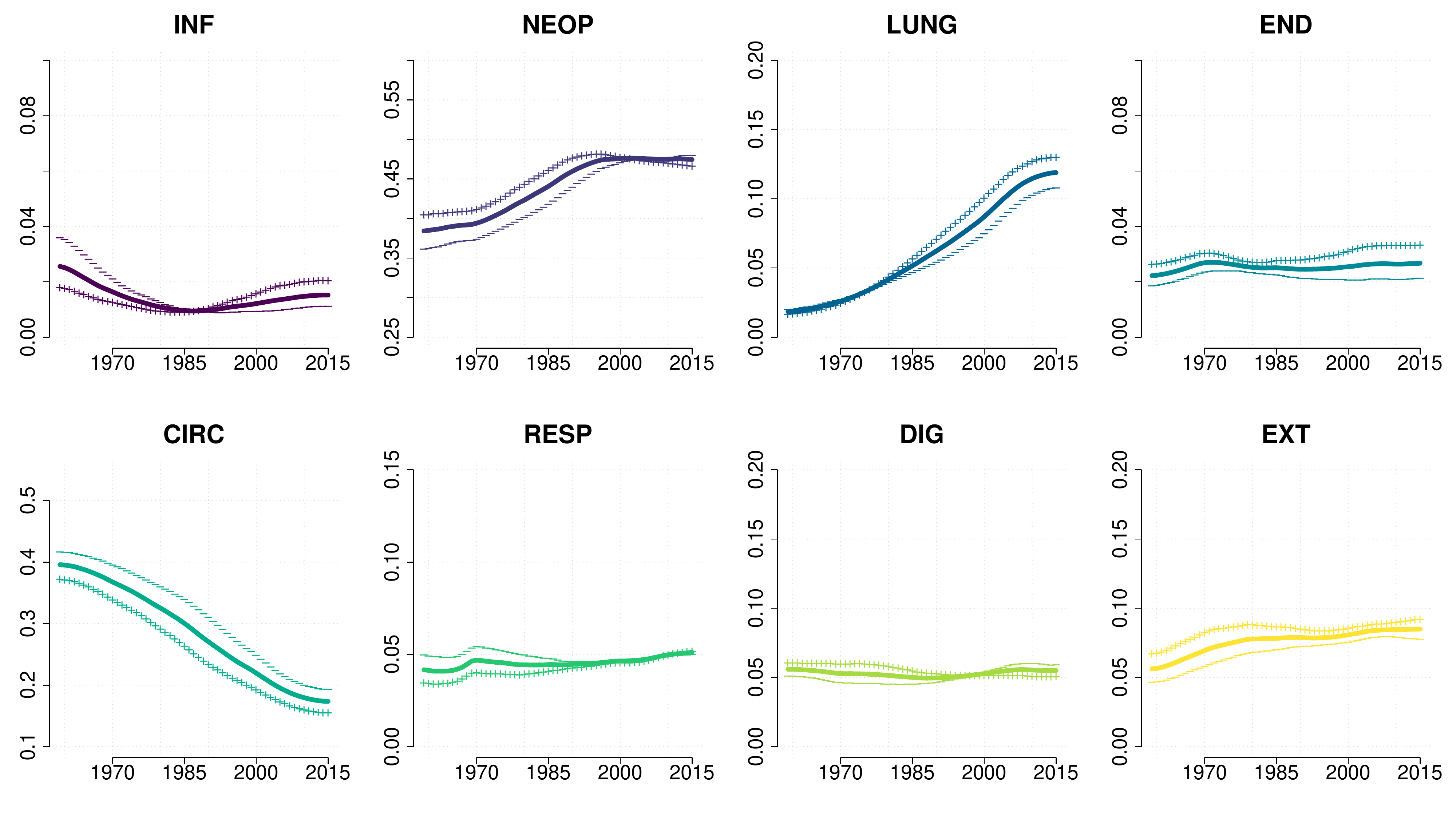}
\caption{Second compositional functional principal component for women population. The continuous line represents the mean function while outer lines represent the mean function +/- the component. }
\label{PC2-women}
\end{figure}

The first compositional functional principal component for women population is shown in figure \ref{PC1-women}. It represents the main mode of variation of the women sample. The eigenvalue associated with this eigenfunction is $\lambda_1 = 14.87$ and the Fraction of Explained Variance (FEV) is $\lambda_1/\sum \lambda_k = 0.326$, thus, similarly to men, this component accounts for about one third of total variability. Surprisingly, this component is related to all causes but circulatory diseases, one of the main causes of death. A {\it level} variability can be found for lung cancer, respiratory and digestive diseases. Similarly for men, {\it local} variability is present in infectious and endocrine diseases, linked to time periods 1959-1985 for the former and 1959-1990 for the latter, but for the women sample we observe {\it local} variability also for external causes, in the interval 1980-2005. Lastly, neoplasms exhibit a variability pattern related to the velocity of the evolution of this cause. Overall, a positive score on this component would lead to a high level of lung cancer and respiratory diseases, a low level of digestive diseases, infectious diseases (especially before 1985), 
\noindent
endocrine diseases (only before 1990) and external causes (between 1980 and 2005) and slow increase of neoplasms. Negative scores characterize an opposite behaviour. The scores obtained by each country on each component for women sample are reported in table \ref{tab:scores}.
The second compositional functional principal component for women population is depicted in figure \ref{PC2-women}. The eigenvalue associated with this eigenfunction is $\lambda_2 = 7.38$ and the FEV is $0.162$. This second component is related to all causes of death. Circulatory and endocrine diseases and external causes show a {\it level} variability. A {\it local} variability can be observed for neoplasms (1959-2000), lung cancer (1980-2015) and respiratory diseases (1959-1995). Lastly, a crossing-mean pattern can be found in infectious and digestive diseases. Overall, a positive score on this component would lead to a low level of circulatory diseases and respiratory diseases (before 1995), a high level of endocrine diseases, external causes, lung cancer (since 1980) and neoplasms (until 2000), and a increase of infectious diseases after 1985.

\begin{figure}[t]
\includegraphics[width=\linewidth]{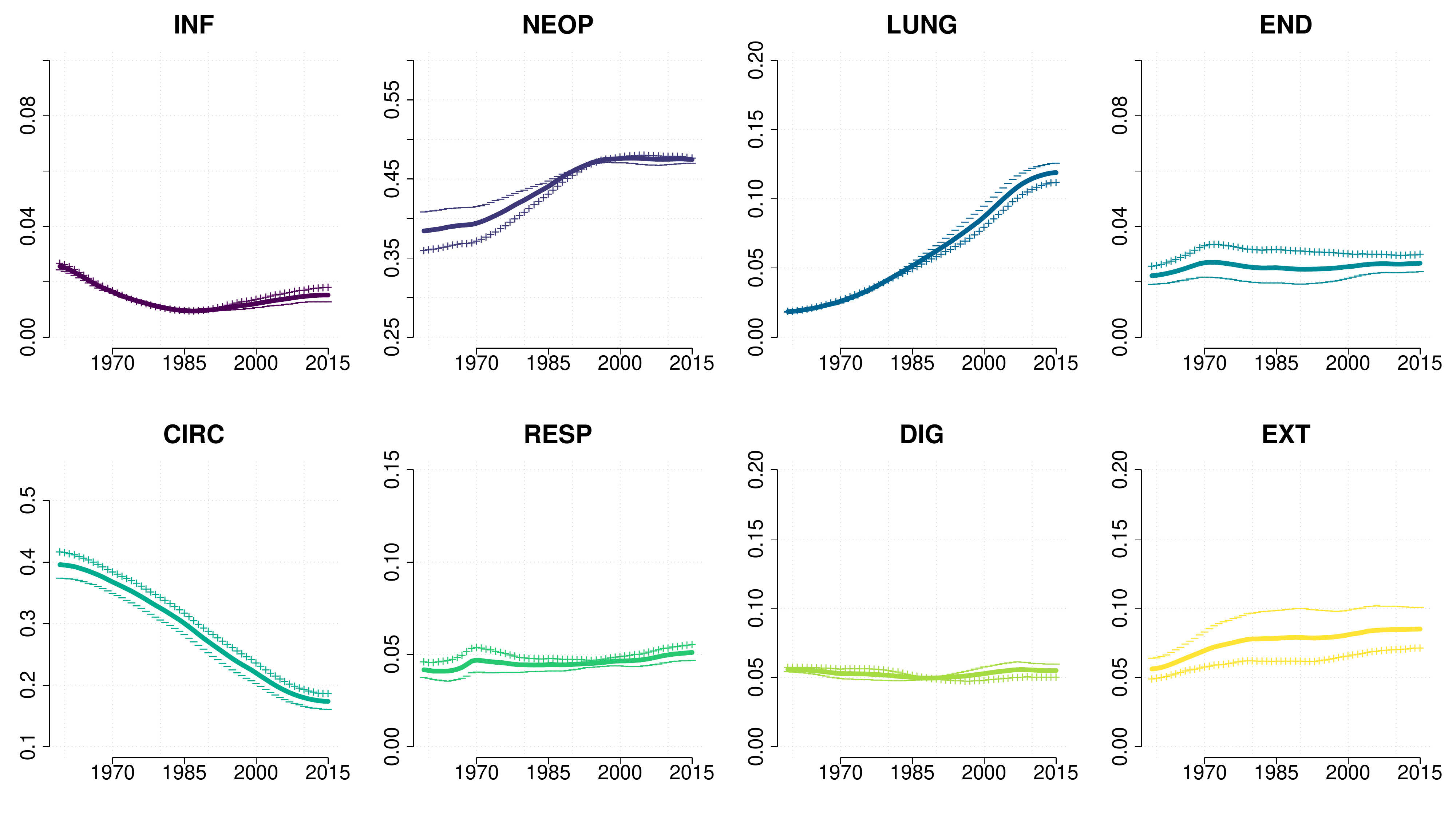}
\caption{Third compositional functional principal component for women population. The continuous line represents the mean function while outer lines represent the mean function +/- the component. }
\label{PC3-women}
\end{figure}

Figure \ref{PC3-women} shows the third compositional functional principal component for women population. The eigenvalue associated with this eigenfunction is $\lambda_3 = 5.94$ and the FEV is $0.13$. 
This component represents a {\it level} variability for endocrine, circulatory and respiratory diseases and for external causes and a {\it local} variability for neoplasms (before 1990) and lung cancer (after 1990).
Overall, a positive score on this component would lead to a high level of endocrine, circulatory and respiratory diseases, a low level of external causes, a low level of neoplasms before 1990 and a low level of lung cancer after 1990. 

\begin{figure}[H]
\includegraphics[width=\linewidth]{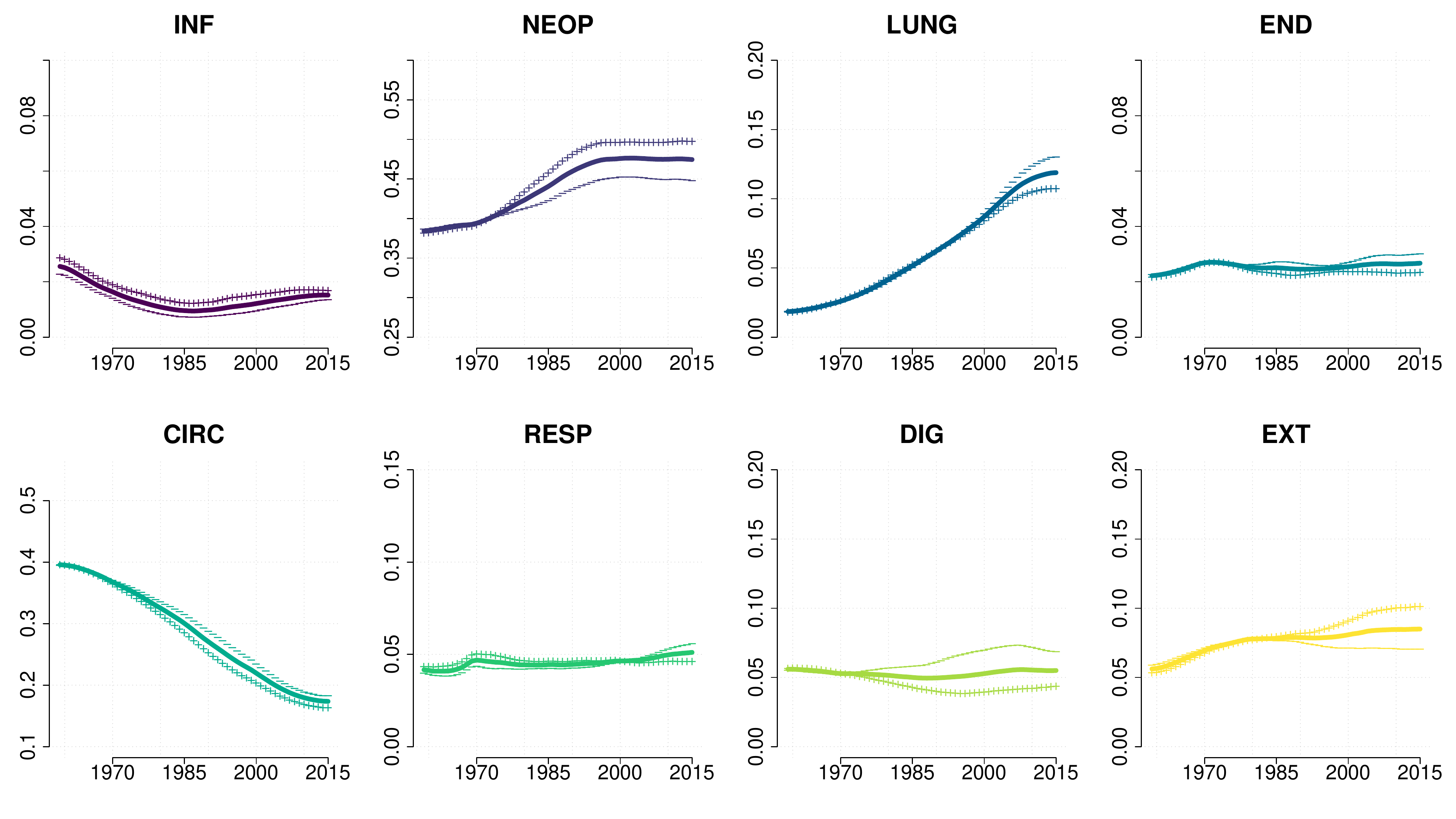}
\caption{Fourth compositional functional principal component for women population. The continuous line represents the mean function while outer lines represent the mean function +/- the component. }
\label{PC4-women}
\end{figure}

\begin{figure}[t]
\includegraphics[width=\linewidth]{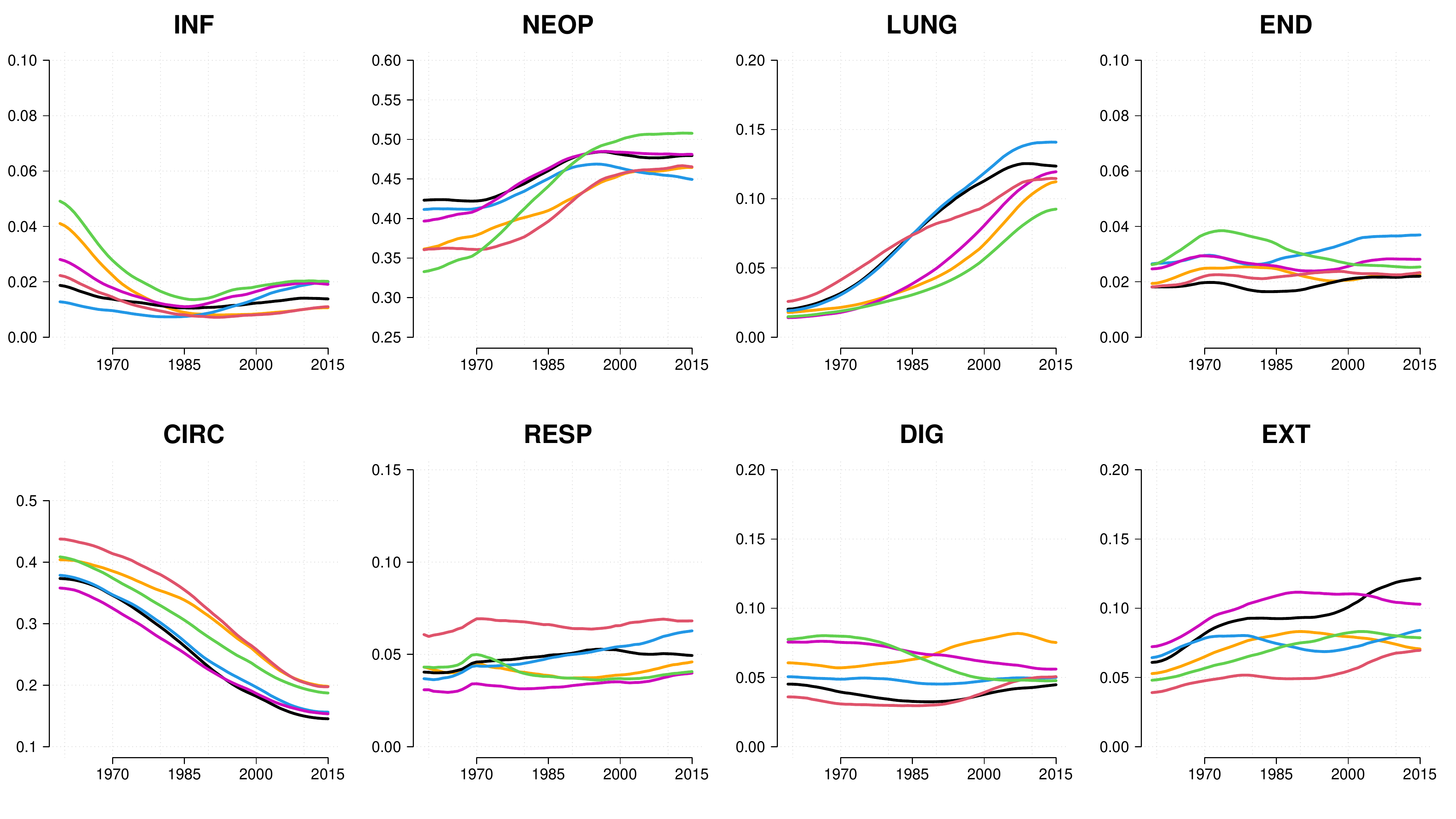}
\caption{Functional centroids of the spectral clustering for women sample. Legend: {\color{Rcol4} \textbf{----}} Cluster 1, {\color{Rcol3} \textbf{----}} Cluster 2, {\color{Rcol6} \textbf{----}} Cluster 3 {\color{ROrange} \textbf{----}} Cluster 4, {\color{Rcol2} \textbf{----}} Cluster 5,  \textbf{----} Cluster 6.}
\label{centroids-women}
\end{figure}

The fourth compositional functional principal component for women population is illustrated in figure \ref{PC4-women}. The eigenvalue associated with this eigenfunction is $\lambda_4 = 5.66$ and the FEV is $0.124$. 
This component represents mainly {\it local} and most recent variability. In particular it is related 
\noindent
to neoplasms after 1980, lung cancer and endocrine diseases after 2000, digestive diseases after 1970 and external causes after 1990. 
Overall, a positive score on this component would lead to 
\noindent
a high level of neoplasms and external causes and a low level of lung cancer and digestive 
diseases, all in the aforementioned periods.
As for men, we applied spectral clustering to the women curves projected on the first $4$ PCs. The percentage of total explained variability is 74.2\%. The algorithm ran $B=1000$ times and we considered different values for $G$. Again, the spectrum of the Laplacian matrix does not support any specific number of clusters, while the silhouette index suggests a value between 6 and 8. For parsimony and interpretability reasons, we choose to use $G=6$ clusters that we present in table \ref{tab:clus-women} (see also figure \ref{scores}). Using (\ref{eq:recon}) with $K=4$ and $\gamma_{1g},\gamma_{2g},\gamma_{3g},\gamma_{4g}$ as the centroids of the spectral clustering output, we reconstructed the compositional functional centroids $\bm{f}_g, g=1,\ldots,G$, depicted in figure \ref{centroids-women}. 

The first group of countries is similar to that of men: it is mainly made up of extra-European countries together with Denmark and Netherlands, with a rising trend of lung cancer and respiratory diseases and endocrine and metabolic diseases, related to increasing prevalence of smoking and obesity among women. The fact that Denmark and Netherlands have been clustered in this group is not surprising, as it is well-known they have undergone a stagnation of life expectancy improvement, and smoking is one of the leading causes of this, as shown also by \cite{Denmark}. The second group is made up by Japan, Italy and Spain, with the lowest level of lung cancer, in contrast to what has been seen for men in the same countries. Indeed, smoking prevalence of women in these countries is much lower than that of men, see \cite{smoke1}. The third group includes France, Belgium and Switzerland; the main feature of this group is the high level of external causes deaths, only recently getting lower than that of Scandinavian countries. The fourth group (which for men includes Hungary, Poland, Greece, and Finland) now includes also Austria and, as for men, it is characterised by high level of digestive system diseases related deaths. It should be noted that also for women this is the least homogeneous group. Austria, in particular, has a peculiar pattern with a recently increasing trend of endocrine and metabolic diseases and a steady level of external causes related deaths. The group with UK and Ireland now includes also New Zealand, instead of Belgium, but the characteristics are similar to the analogous group for men: high respiratory and circulatory diseases and increasing digestive system diseases related deaths. Finally, the Scandinavian group: as for men, the main feature is the high mortality for external causes but, in contrast with men, this group is also characterized by a high and increasing relevance of lung cancer related mortality.

\section{Conclusions}
\label{sec:conc}
We have proposed to combine compositional data analysis (CDA) and functional data analysis (FDA) to describe the global trend of cause--specific mortality of several countries. CDA allows us to analyze cause--specific mortality rates (CSMRs) taking properly into account their competing---risk nature and FDA is then applied to compositional data so that their trends can be decomposed by means of Functional principal components analysis and countries can be clustered by means of PCs we have retrieved. In this way we are able to make a descriptive but comprehensive analysis of trends of CSMRs. 
Results give us many insights of the ongoing trend that the considered populations are undergoing in terms of composition of causes of deaths and, while many of them come as no surprise, in some cases we find some evidence that has not been highlighted by past literature. We should bear in mind that our analysis is focused on ages 40--64, so all considerations should be applied to mid-life mortality only.\\
The first finding is that clusters for men and women are very similar, albeit some differences can be found: the main one is that for women we have one cluster more, which is basically the result of splitting the cluster 2 into two. However, even though the clustering results are quite similar for men and women, we can easily notice that evolution of mortality composition is quite different: in all countries, prevalence of lung cancer among men has stopped increasing between 1990 and 2000, while for women it kept on rising until the very recent years, when first signs of plateauing can be seen. Moreover for some countries we can see a high disparity between men and women (e.g., Italy, Japan and Spain are in the cluster with lowest level of lung cancer for women and in that with the highest for men, while for cluster including Nordic countries is the other way round) while for others (e.g., East European countries) women catch up with men.
Relevance of digestive system and metabolic, endocrine and nutritional diseases related deaths is increasing especially for men, suggesting that poor dieting and alcohol consumption are increasingly impacting on men's health. For women a rising concern is lung cancer and respiratory diseases, especially in Nordic countries. Interestingly some countries have a peculiar pattern that is difficult to group with others. Finland, in particular, with extremely high external causes and digestive system diseases related deaths, shows a composition of causes of death that is difficult to classify. Such peculiarity is confirmed by the relatively low life expectancy with respect to other Scandinavian countries.\\
Although FDA and CDA approaches are helpful in explaining cause--specific mortality trends in a comprehensive way, both of them come with a specific limitation. The limitation of FDA is that it allows a descriptive analysis through functional principal components and cluster analysis, but it can not be used in a straightforward way to forecast the future trends. A forecast application that takes into account the compositional aspect but not the functional one has been implemented by \cite{Oeppen1}. The limitation of CDA is that it focus on cause--specific rates, but the overall trend is not considered. However, the clusters we identified largely capture differences in overall mortality rate trends of countries.\\
On the other hand, such a combination of CDA and FDA can be helpful in other applications: the same analysis can be applied to older ages (65+), where, following \cite{Horiuchi2003}, we can expect to find higher prevalence of infectious diseases, mental disorders and cerebrovascular diseases. Another possible application, remaining in the demographic field but turning to a different aspect, could be to describe trends of parity--specific fertility rates, which can be seen as a composition of the overall fertility, although it could be hard to find a sufficient number of countries with parity--specific fertility data for a long enough time window.

\section*{Acknowledgements}

We acknowledge the support from MIUR--PRIN 2017 project number 20177BR-JXS.

\bibliographystyle{rss}
\bibliography{references}

\begin{thebibliography}{20}
\expandafter\ifx\csname natexlab\endcsname\relax\def\natexlab#1{#1}\fi
\expandafter\ifx\csname url\endcsname\relax
  \def\url#1{\texttt{#1}}\fi
\expandafter\ifx\csname urlprefix\endcsname\relax\def\urlprefix{URL: }\fi

\bibitem[{Aitchinson(1986)}]{Aitchinson}
Aitchinson, J. (1986) \textit{The Statistical Analysis of Compositional Data}.
\newblock Chapman \& Hall.

\bibitem[{Brennan and Bray(2002)}]{smoke1}
Brennan, P. and Bray, I. (2002) Recent trends and future directions for lung
  cancer mortality in europe.
\newblock \textit{British Journal of Cancer}, \textbf{87}, 43--48.

\bibitem[{Canudas-Romo et~al.(2020)Canudas-Romo, Aldair and
  Mazzuco}]{TCAL_2020}
Canudas-Romo, V., Aldair, T. and Mazzuco, S. (2020) Cause of death
  decomposition of cohort survival comparisons.
\newblock \textit{International Journal of Epidemiology}.
\newblock \urlprefix\url{https://doi.org/10.1093/ije/dyz276}.
\newblock Dyz276.

\bibitem[{Egozcue and Pawlowsky-Glahn(2011)}]{pawlowsky}
Egozcue, J.~J. and Pawlowsky-Glahn, V. (2011) \textit{Compositional Data
  Analysis: Theory and Applications}.
\newblock John Wiley \& Sons, Ltd.

\bibitem[{Fryar et~al.(2014)Fryar, Carroll and Ogden}]{USA_obesity}
Fryar, C.~D., Carroll, M.~D. and Ogden, C.~L. (2014) Prevalence of overweight,
  obesity, and extreme obesity among adults: United states, 1960--1962 through
  2011--2012.
\newblock \textit{Tech. rep.}, National Center for Health Statistics.

\bibitem[{Horiuchi et~al.(2003)Horiuchi, Finch, Meslé and
  Vallin}]{Horiuchi2003}
Horiuchi, S., Finch, C.~E., Meslé, F. and Vallin, J. (2003) {Differential
  Patterns of Age-Related Mortality Increase in Middle Age and Old Age}.
\newblock \textit{The Journals of Gerontology: Series A}, \textbf{58},
  B495--B507.

\bibitem[{Hron et~al.(2016)Hron, Menafoglio, Templ, Hr\r{u}zov\'a and
  Filzmoser}]{menafoglio}
Hron, K., Menafoglio, A., Templ, M., Hr\r{u}zov\'a, K. and Filzmoser, P. (2016)
  Simplicial principal component analysis for density functions in bayes
  spaces.
\newblock \textit{Computational Statistics \& Data Analysis}, \textbf{94}, 330
  -- 350.

\bibitem[{Izenman(2008)}]{multivariate}
Izenman, A.~J. (2008) \textit{Modern Multivariate Statistical Techniques:
  Regression, Classification, and Manifold Learning}.
\newblock Springer.

\bibitem[{Kj{\ae}rgaard et~al.(2019)Kj{\ae}rgaard, Ergemen, Kallestrup-Lamb,
  Oeppen and Lindahl-Jacobsen}]{Oeppen1}
Kj{\ae}rgaard, S., Ergemen, Y.~E., Kallestrup-Lamb, M., Oeppen, J. and
  Lindahl-Jacobsen, R. (2019) Forecasting causes of death using compositional
  data analysis: the case of cancer deaths.
\newblock \textit{Journal of the Royal Statistical Society: Series C (Applied
  Statistics)}, \textbf{68}, 1351--1370.

\bibitem[{Kraus(2015)}]{kraus}
Kraus, D. (2015) Components and completion of partially observed functional
  data.
\newblock \textit{Journal of the Royal Statistical Society. Series B
  (Statistical Methodology)}, \textbf{77}, 777--801.

\bibitem[{Leng and Müller(2006)}]{muller2}
Leng, X. and Müller, H.-G. (2006) Classification using functional data
  analysis for temporal gene expression data.
\newblock \textit{Bioinformatics}, \textbf{22}, 68--76.

\bibitem[{Lindahl-Jacobsen et~al.(2016)Lindahl-Jacobsen, Rau, Jeune,
  Canudas-Romo, Lenart, Christensen and Vaupel}]{Denmark}
Lindahl-Jacobsen, R., Rau, R., Jeune, B., Canudas-Romo, V., Lenart, A.,
  Christensen, K. and Vaupel, J.~W. (2016) Rise, stagnation, and rise of danish
  women{\textquoteright}s life expectancy.
\newblock \textit{Proceedings of the National Academy of Sciences},
  \textbf{113}, 4015--4020.
\newblock \urlprefix\url{https://www.pnas.org/content/113/15/4015}.

\bibitem[{von Luxburg(2007)}]{spectral}
von Luxburg, U. (2007) A tutorial on spectral clustering.
\newblock \textit{Statistics \& Computing}, \textbf{17}, 395 -- 416.

\bibitem[{Oeppen(2008)}]{Oeppen2}
Oeppen, J. (2008) Coherent forecasting of multiple-decrement life tables: A
  test using japanese cause of death data.
\newblock \textit{Tech. rep.}, Catalonia, Spain: Departament d'Inform\`atica i
  Matem\`atica Aplicada, Universitat de Girona.

\bibitem[{Petersen and Müller(2016)}]{muller}
Petersen, A. and Müller, H.-G. (2016) Functional data analysis for density
  functions by transformation to a hilbert space.
\newblock \textit{Ann. Statist.}, \textbf{44}, 183--218.

\bibitem[{Preston et~al.(2001)Preston, Heuveline and Guillot}]{Preston2001}
Preston, S.~H., Heuveline, P. and Guillot, M. (2001) \textit{Demography.
  Measuring and Modeling Population Processes}.
\newblock Blackwell Publishing.

\bibitem[{Ramsay and Silverman(2005)}]{FDA1}
Ramsay, J.~O. and Silverman, B.~W. (2005) \textit{Functional data analysis},
  vol. 2nd Ed.
\newblock New York: Springer.

\bibitem[{Sangalli et~al.(2009)Sangalli, Secchi, Vantini and
  Veneziani}]{sangalli}
Sangalli, L.~M., Secchi, P., Vantini, S. and Veneziani, A. (2009) A case study
  in exploratory functional data analysis: Geometrical features of the internal
  carotid artery.
\newblock \textit{Journal of the American Statistical Association},
  \textbf{104}, 37--48.

\bibitem[{Woolf and Schoomaker(2019)}]{Woolf2019}
Woolf, S.~H. and Schoomaker, H. (2019) {Life Expectancy and Mortality Rates in
  the United States, 1959-2017}.
\newblock \textit{Journal of the American Medical Association}, \textbf{322},
  1996--2016.

\bibitem[{{World Health Organization}(2019)}]{WHO}
{World Health Organization} (2019) Mortality database.
\newblock available at
  \url{https://apps.who.int/healthinfo/statistics/mortality/causeofdeath_query/start.php}.

\end{thebibliography}
\end{document}